\documentstyle[amsmath,amssymb,aps,twocolumn,graphicx,floats]{revtex}

\title{AC electric and magnetic responses of non-connected
Aharonov-Bohm rings}

\author{R. Deblock$^1$, Y. Noat$^2$, B. Reulet$^1$, H. Bouchiat$^1$,
  and D. Mailly$^3$}

\address{ $^1$ Laboratoire de Physique des Solides, Associ\'e au CNRS,
B\^atiment 510, Universit\'e Paris-Sud, 91405, Orsay, France.\\ $^2$
Kamerlingh Onnes Laboratory, Niels Bohrweg 5, Leiden, Netherland.\\
$^3$ CNRS Laboratoire de Microstructure et Micro\'electronique, 196
avenue Ravera, 92220, Bagneux, France.}

\begin{document}
\maketitle
\begin{abstract}
Signature of phase coherence on the electric and magnetic response of
$10^5$ non-connected Aharonov-Bohm rings is measured by a resonant method at
350 MHz between 20 mK and 500 mK. The rings are etched in a
GaAs-AlGaAs heterojunction.  Both quantities exhibit an oscillating
behaviour with a periodicity consistent with half a flux quantum
$\Phi_0/2=h/2e$ in a ring. We find that electric screening is enhanced
when time reversal symmetry is broken by magnetic field, leading to a
\textit{positive} magnetopolarisability, in agreement with theoretical
predictions for isolated rings at finite frequency. Temperature and
electronic density dependence are investigated. The dissipative part
of the electric response, the electric absorption, is also measured
and leads to a \textit{negative} magnetoconductance.  The magnetic
orbital response of the very same rings is also investigated. It is
consistent with \textit{diamagnetic} persistent currents of 0.25 nA.
This magnetic response is an order of magnitude smaller than the
electric one, in qualitative agreement with theoretical expectations.
\end{abstract}

\section{Introduction}
At mesoscopic scale and at low temperature, electrons in metallic
samples keep their phase coherence on a length $L_{\Phi}$ which is
bigger than the sample size. Transport and thermodynamic
properties of the system are then sensitive to interference between
electronic wave functions and present spectacular signatures of this
phase coherence. To study these effects the ring geometry is
especially suitable. Indeed in the presence of a magnetic flux
$\Phi$ through the ring the periodic boundary conditions for
electronic wavefunctions acquire a phase factor $2 \pi \Phi/\Phi_0$
with $\Phi_0=h/e$ the flux quantum \cite{intro}.  As a result, the
magnetoconductance of a phase coherent ring exhibits quantum
oscillations which periodicity corresponds to one flux quantum through
the area of the sample \cite{washburn86}. The phase of the first
harmonics of these oscillations is sample specific so that these
harmonics do not survive ensemble average. In contrast the second harmonics
have a contribution which resists this averaging. This results from
the interference between time reversed paths around the ring (weak
localisation contribution). These $\Phi_0/2$ periodic oscillations
were observed in transport measurements on long cylinders or connected
arrays of rings \cite{sharvin81,aronov87}. Their sign corresponds to a
positive magnetoconductance in zero field. In the case of singly
connected geometries, like full disks, the signature of weak
localization consists in a single peak of positive magnetoconductance
which width corresponds to $\phi_0/2$ through the sample
\cite{chang94,lee97}.

Magneto-transport experiments on connected systems constitute a very
sensitive and powerful probe for the investigation of sample specific
signatures of quantum transport. However, because of strong coupling
between the system and the measuring device, quantum corrections
represent a small fraction of the conductance (of the order of $1/g$
where $g$ is the dimensionless conductance expressed in $e^2/h$ units)
which is still dominated by the classical Drude response in the
diffusive regime $g \gg 1$. There exists a number of experiments which
can address some of the electronic properties of mesoscopic samples
without coupling to macroscopic wires. This is the case of ac
conductance experiments where Aharonov Bohm rings are coupled to an
electromagnetic field.  In contrast with the connected case, the
response of an isolated system can be dominated by quantum
effects. Moreover, the quasi discrete nature of the energy spectrum
and the sensitivity to the statistical ensemble (Canonical or
Grand-Canonical) are new features of isolated mesoscopic systems. In
particular it has been emphasized that the average absorbtion of
isolated mesoscopic systems is determined by the energy level
statistics and its sensitivity to time reversal symmetry breaking by a
magnetic field.

The first experiments done in this spirit were performed by coupling
an array of disconnected GaAs/GaAlAs rings to a strip-line
superconducting resonator \cite{reulet95}. In such a geometry the
rings experience both ac magnetic and electric field. The magnetic
response of the rings i.e. their orbital magnetism is directly related
to persistent currents in the zero frequency limit
\cite{buttiker83,eckern95,chandrasekhar91,jariwala01,mailly93,levy90}.
On the other hand the electric response of isolated metallic sample is
related to the screening of the electric field inside the metal.  The
induced charge displacement is at the origin of the polarisability
$\alpha$, defined as the ratio between the induced electric dipole
$\mathbf{d}$ and the applied electric field $\mathbf{E}$ ($\mathbf{d}=
\alpha \mathbf{E}$). The polarisability is known to be essentially
determined by the geometry of the sample with correction of the order
of $\lambda_s/L$, with $\lambda_s$ the screening length and $L$ the
typical size of the system \cite{rice}.  It has been recently
predicted that this quantity is sensitive to phase coherence around
the ring \cite{efetov96,noat96,blanter98} and is thus expected to
present flux oscillations.  The electric contribution can be in the
particular case of GaAs rings of the same order of magnitude as the
magnetic response \cite{noat98SM}.

To be able to distinguish between the two types of response we have
designed a superconducting LC resonator which capacitive part and
inductive part are physically separated. In this paper we present
measurements of both magnetic and electric response of Aharonov-Bohm
rings. Note that these experiments are done on the very same array of 
rings for both types of response, giving us the opportunity to compare
them. Preliminary account of measurement of the electric response was
given in reference \cite{deblock00}.

The paper is organised in the following way. Section
\ref{experimentalsetup} gives a detailed presentation of the sample,
an array of Aharonov-Bohm rings, and the resonating technic used to
measure the magnetic and electric response.  Results on the
non-dissipative part of the flux dependent electric response are 
presented in section \ref{magnetopolarisabilty}. A comparison with
theoretical predictions is given, including frequency
dependence. Temperature and electronic density dependence are also
investigated. The next section focuses on the dissipative part of the
magnetopolarisability of the rings.  Theoretical results for this
quantity are derived and compared to the experiment.  The section
\ref{magneticresponse} is devoted to the measurements of the magnetic
response of the same rings. Despite the fact that the signal is then
smaller, the magnetic response of the rings is detected and compared
to predictions on averaged persistent currents. We conclude by a
comparison between the magnetic and electric response.

\section{Experimental setup}
\label{experimentalsetup}

\subsection{The sample}

\subsubsection{The rings}
\label{sec:rings} 

\begin{table*}[tb]
\begin{center}
\begin{tabular}{|c|c|c|}
 & & \\ Nominal density $n$& &$3\,10^{11}$ cm$^{-2}$\\ Thomas Fermi
screening length $\lambda_s$ \cite{ando} &$\lambda_s = (\pi/2) \, 4 \pi \epsilon_0
\epsilon_r \hbar^2 /(m e^2)$ &16 nm\\ Perimeter $L$ & & 5.2
$\mu$m \\ Etched width & &0.5 $\mu$m \\ Effective width $W$ \cite{wl}
& & 0.2 $\mu$m\\ Phase coherence length $L_{\Phi}$ \cite{wl} & &6.5
$\mu$m\\ Mean free path $l_e$ \cite{wl} & & 3 $\mu$m\\ Diffusion
coefficient $D$ & $D=v_F l_e / 2$ & 0.335 m$^2$.s$^{-1}$\\ Mean level
spacing $\Delta$ & $\Delta = h^2 / (2 \pi m W L)$ & 80 mK or 1.66 GHz
\\ Thouless energy $E_c$ & $E_c = h D / L^2$ & 450 mK or 9.34 GHz \\
dimensionless conductance $g$ & $g=E_c/\Delta$ & 5.6 \\ & & \\
\end{tabular}
\end{center}
\caption{Characteristics of the rings after illumination.}
\label{tab:CharacteristicsOfTheRings} 
\end{table*}

\begin{figure}[tb]
    \begin{center}
        \includegraphics[width=8cm]{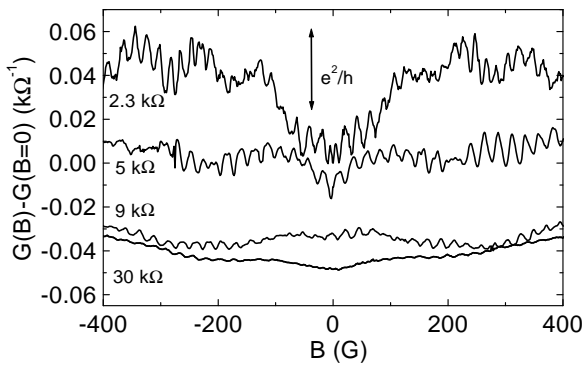}
    \end{center}
    \caption{Conductance of an Aharonov-Bohm ring at different illuminations. 
    At zero illumination the conductance is zero. With illumination
    the resistance decreases. The curves are shifted for clarity.}
    \label{fig:abillum}
\end{figure}
\begin{figure}[tb]
        \begin{center}
         \includegraphics[width=8cm]{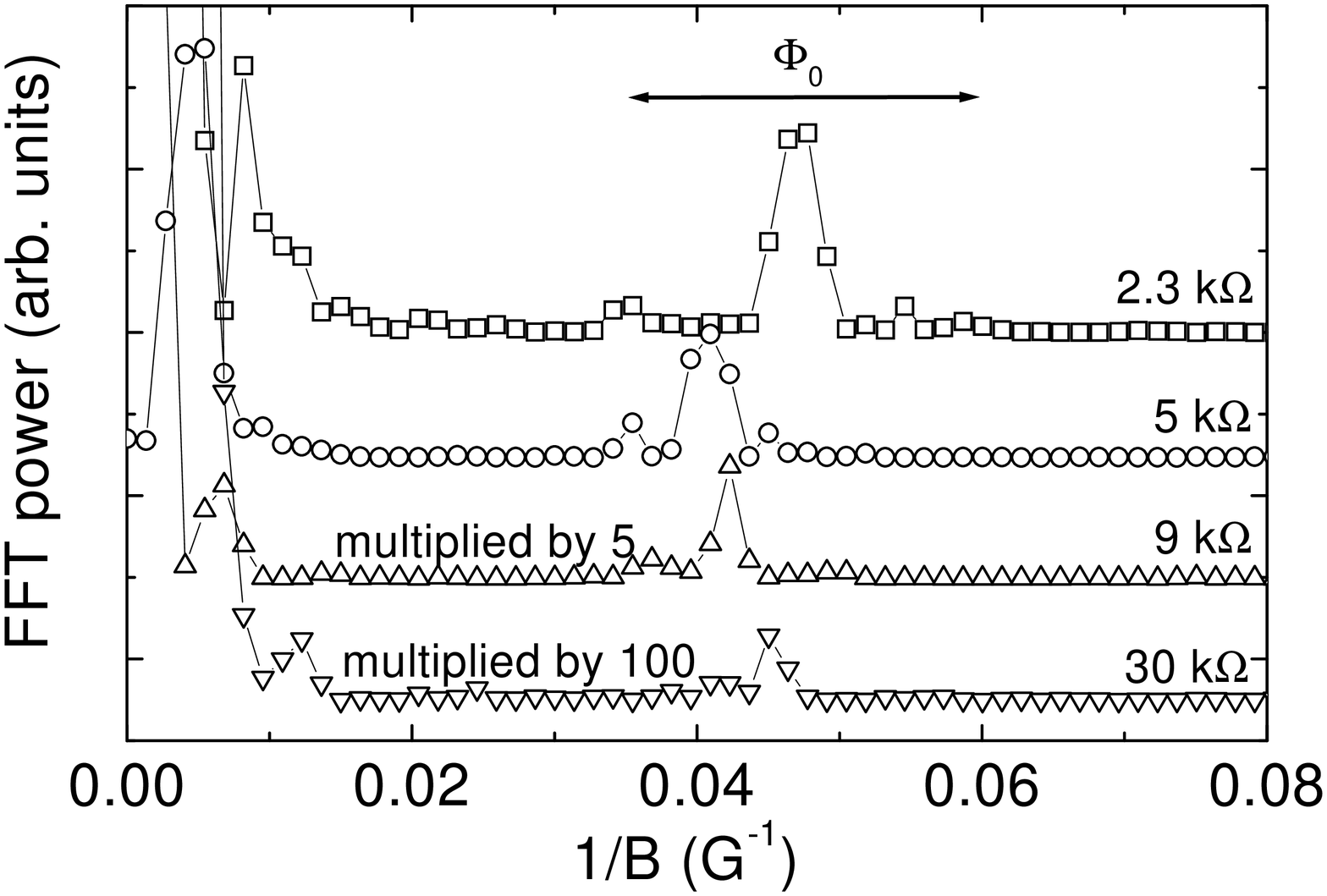}
        \end{center}
        \caption{Fourier transform of the magnetoconductance of an 
        Aharonov Bohm ring at different illumination. The shape and
        amplitude of the $\Phi_0$ peak is strongly dependent on
        illumination. The curves are shifted for clarity.}
        \label{fig:FFTabillum}
\end{figure}

We have studied the electric and magnetic susceptibilities of isolated
Aharonov-Bohm rings. Our system is an array of $10^5$ 2D rings etched
by reactive ion etching in a high mobility AlGaAs/GaAs
heterojunction. The characteristics of these rings are given in 
table \ref{tab:CharacteristicsOfTheRings}.  They are ballistic in the
transverse direction and diffusive longitudinally ($l_e < L$ and $l_e
\gg W$). It is important to perform a deep etching of the
heterojunction (down to GaAs) in order to minimize high frequency
losses, which have been observed to be important in etched AlGaAs. Because
of etching the electronic density is strongly depressed. However we
are able to recover the nominal density of the heterojunction by
illuminating the rings with a infrared diode placed close to the
sample in the dilution refrigerator. For each illumination a current
of $10 \mu A$ is run through the diode during several minutes. 
Measurements are done at least one hour after the illumination in
order to ensure good stability of the sample. An upper value of the
estimated illumination power coupled to the sample is 600 photons/s
with a wavelength of 766~nm. With this setup we are thus able to
perform measurements at different electronic density. The
control on the density is rather qualitative because of the difficulty
to calibrate the illumination procedure.  We have checked the effect
of illumination on a connected Aharonov-Bohm (AB) ring (figure
\ref{fig:abillum}). At zero illumination time the conductance of the
ring is zero. On such a sample we can follow the AB oscillations when
the resistance decreases by more than an order of magnitude with
illumination.  As a consequence a clear effect of illuminating the
ring is to increase its conductance. The Fourier transform of the 
resistance of the ring is shown on figure
\ref{fig:FFTabillum}. We see for each illumination an oscillation
whose periodicity is consistent with a flux quantum $\Phi_0$ in the
area of the ring. However the Fourier transform shows that the peak
corresponding to this periodicity changes with illumination both in
shape and in amplitude. The amplitude increases with illumination
due to the increase of AB oscillations. The fact that the shape, and
in particular the width, of the peak changes with illumination is an
indication that the width of the ring increases with illumination
time. To be more precise the width of the rings is multiplied by a
factor 2 between the first and last curve of figure
\ref{fig:FFTabillum}. Note that the increase of the electronic 
density has also been shown to induce an increase of the electronic 
mobility \cite{sajoto90}.

\begin{figure}[tb]
        \begin{center}
                \includegraphics[width=8cm]{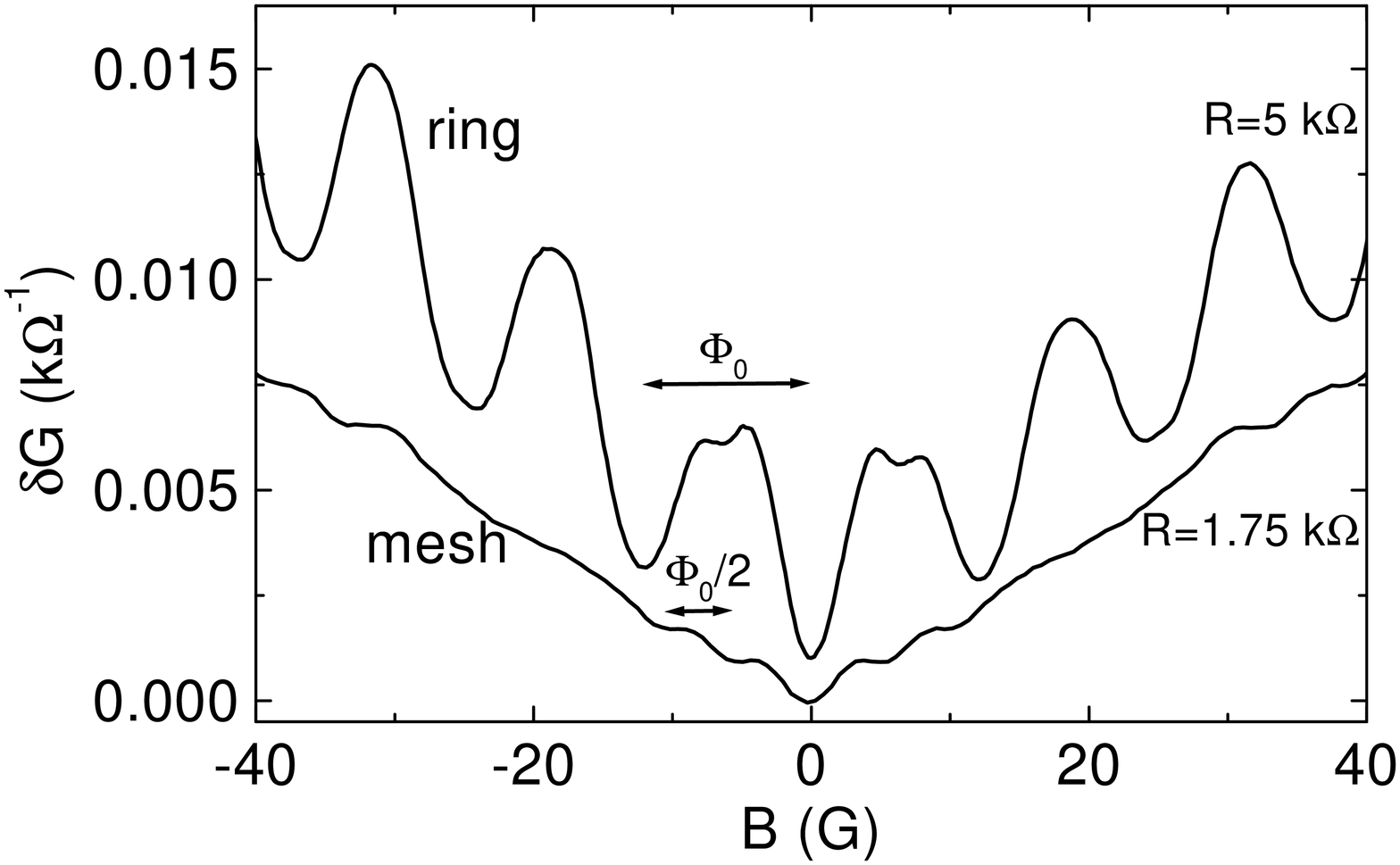}
        \end{center}
        \caption{Magnetoconductance of a ring and a mesh. The $\Phi_0$ signal
        disappears with ensemble average, so that in the mesh only the
        $\Phi_0/2$ component remains. Note the triangular shape of the
        magnetoconductance on the mesh.  The curves are shifted for
        clarity.}
        \label{fig:angr}
\end{figure}

In order to study the disorder average we have measured conductance of
a single ring and a mesh, representing a two-dimensional square array,
etched in the same type of heterojunction than the rings. The
magnetoconductance is shown on figure \ref{fig:angr}. As expected the
AB effect disappears under ensemble averaging. The $\Phi_0/2$
oscillations on the other hand remains on the mesh. In this case the
triangular shape of the magnetoconductance is attributed to weak
localisation in the wire of the mesh.

\subsubsection{Superconducting micro-resonator}
\begin{figure}[tb]
    \begin{center}
        \includegraphics[width=8cm]{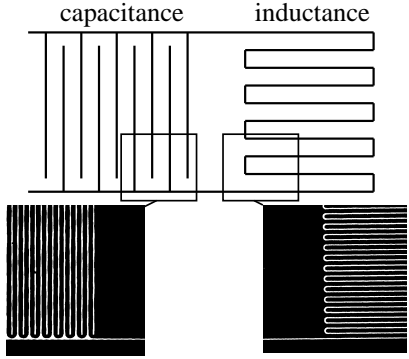}
    \end{center}
    \caption{Schematic drawing of the resonator and optical 
    photographes of part of it. Note that the inductance (meander
    line) is physically separated from the capacitance (comb-like
    structure).}
    \label{fig:resonator}
\end{figure}
To measure the electric or magnetic response of the rings we couple
them to a superconducting micro-resonator and detect the changes in
its properties.  This resonator is made by optical lithography. It
consists of a niobium strip-line deposited on a sapphire
substrate. This substrate has been preferred to silicon or GaAs
because it induces the weakest temperature dependence of the resonance
frequency and gives the best quality factor due to the quality of the
niobium layer on sapphire. A schematic drawing is given in figure
\ref{fig:resonator}. The width of the wire constituting the resonator
is 2 $\mu$m, the thickness 1 $\mu$m and the spacing between two
adjacent wires is 4 $\mu$m. The total length of the capacitance or the
inductance is 10 or 20 cm.  In this kind of resonator the inductance
is physically separated from the capacitance by a distance of 300
$\mu$m, allowing to submit the sample only to an electric field (or to
a magnetic field) to measure its electric (or magnetic) response.
This separation between magnetic and electric response has been
checked by deposition of a paramagnetic system (DPPH) alternatively on
the capacitive and inductive part of the resonator. A magnetic spin
resonance signal was only observed when DPPH was on the inductive
part. The resonance frequency of the bare resonator varies between 200
MHz and 400 MHz depending on the geometry.  Its quality factor is
10000 at 4.2 K and 200000 at 20 mK.  The resonator can be modelled by
an LC circuit of resistance $r$, inductance $\mathcal{L}$, capacitance
$\mathcal{C}$, whose resonance frequency is $f_0=1/ 2 \pi
\sqrt{\mathcal{L} \mathcal{C}}$ and quality factor $Q = {\mathcal{L}}
\omega_0 /r$.  From the value of the higher resonance frequencies of
the resonator we have estimated that the residual capacitance of the
meander line is at least 10 times smaller than $\mathcal{C}$.  Due to
the Meissner effect, the dc field just above the resonator is strongly
inhomogeneous. In order to minimize this effect, we have inserted a
thin, $1 \mu$m-thick, mylar film between the detector and the rings. This
reduces the field inhomogeneity to about $10$\%, which is of the order
of fluctuations in the lithography.

\subsubsection{Electric coupling between the rings and the resonator}
\label{secelectriccoupling} 
In order to measure their electric response the rings are placed on
top of the capacitance of the resonator. Note that with this procedure
the rings are not well aligned with the resonator so that they do not
have the same coupling with the capacitance. This is not a problem as
soon as only linear response is investigated. We note
$\alpha(\omega)=\alpha^{'}(\omega) - i \alpha^{''} (\omega)$ the
polarisability averaged over disorder of a ring at the frequency
$\omega$. The impedance of the capacitance $\mathcal{C}$ slightly
modified by the rings reads~:
\begin{eqnarray*}
    Z(\omega) &=& \frac{1}{i {\mathcal{C}} \omega (1 + N k_e
    \alpha(\omega))} \\ &\approx& \frac{1}{i {\mathcal{C}} \omega} (1
    - N k_e \alpha^{'}(\omega) + i N k_e \alpha^{''}(\omega))
\end{eqnarray*}
In this expression $N$ is the number of rings coupled to the
capacitance, $k_e$ is an averaged coefficient measuring the dielectric
coupling between one ring and the capacitance. The capacitance with
the rings is equivalent to a capacitance ${\mathcal{C}} (1 + N k_e
\alpha^{'}(\omega))$ in series with a resistance $N k_e
\alpha^{''}(\omega)/{\mathcal{C}} \omega$. Hence~:
\begin{equation}
    \frac{\delta \mathcal{C}}{\mathcal{C}}= N k_e \alpha^{'}(\omega)
\end{equation}
The frequency shift due to the rings is then~:
\begin{equation}
    \label{frequencyshift}
    \frac{\delta f}{f_0}= -\frac{1}{2} N k_e \alpha^{'}(\omega_0)
\end{equation}
The quality factor is determined by~:
\begin{equation}
    \frac{1}{Q}=\frac{r+ \frac{N k_e
    \alpha^{''}(\omega_0)}{{\mathcal{C}} \omega_0}}{{\mathcal{L}}
    (\omega_0 + \delta \omega)}
\end{equation}
so that, with ${\mathcal{L}} {\mathcal{C}} \omega_0^2 = 1$ at
resonance~:
\begin{equation}
    \label{Qshift}
    \delta (\frac{1}{Q}) = N k_e \alpha^{''}(\omega_0) - \frac{1}{Q}
    \frac{1}{2} k_e N \alpha^{'}(\omega_0) \approx N k_e
    \alpha^{''}(\omega_0)
\end{equation}
provided that $Q \gg 1$.

The electric coupling coefficient is estimated in appendix
\ref{electric_coupling}. Knowing this value and the number of rings
coupled to the resonator, it is possible to evaluate quantitatively
the polarisability of the rings by measuring the resonance frequency
shift (equation \ref{frequencyshift}) and the variation of the quality
factor (equation \ref{Qshift}).

\subsubsection{Magnetic coupling with the resonator}
When the rings are placed on top of the inductance $\mathcal{L}$ of
the resonator, this inductance is shifted because of their magnetic
response $\chi(\omega) = \chi^{'}(\omega) - i \chi^{''}(\omega)$
according to~:
\begin{equation}
    \frac{\delta \mathcal{L}}{\mathcal{L}} = N k_m \chi
\end{equation}
with $N$ the number of rings coupled to the resonator, $k_m$ the
magnetic coupling coefficient between one ring and the inductance,
which has the dimension of the inverse of a volume.  Note that,
properly defined, the coupling coefficient $k_m$ is of the same order
of magnitude than $k_e$.  More precisely the estimation of $k_e$ and
$k_m$ done in appendix leads to $k_m \approx \epsilon_0 \epsilon_r
k_e$, as expected from reference \cite{noat98SM}.  Following the same
reasoning than for the electric coupling, the properties of the
resonator are modified according to~:
\begin{eqnarray}
    \label{dfmagnetic}
    \frac{\delta f}{f} &=& -\frac{1}{2} N k_m \chi^{'}\\ \delta
    (\frac{1}{Q}) &=& N k_m \chi^{''}
\end{eqnarray}

From previous equations it is in principle possible to measure the
absolute value of $\alpha$ or $\chi$. However when a GaAs sample is 
on the inductive or capacitive part of the resonator the modification
of the resonance is dominated by the influence of the substrate.
As a consequence it is very difficult to have an accurate absolute
measurement. Nevertheless relative measurements are possible so that
the variation of the electric or magnetic response with magnetic field
can be detected in a reliable way.

\subsection{Measurement of the resonance frequency and the quality factor}
\begin{figure}[tb]
    \begin{center}
        \includegraphics[width=8.5cm]{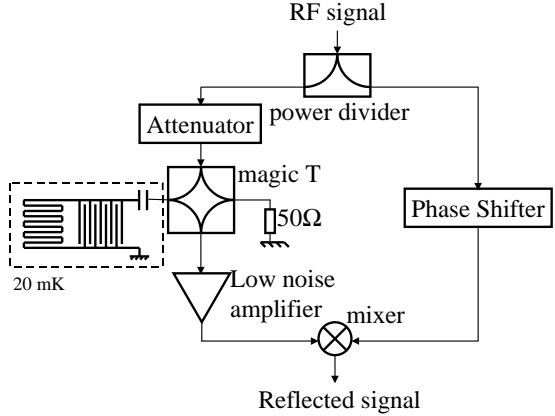}
    \end{center}
    \caption{Rf circuit for measuring the reflected signal from the resonator}
    \label{fig:reflectedsignal}
\end{figure}
\begin{figure}[tb]
    \begin{center}
        \includegraphics[width=8.5cm]{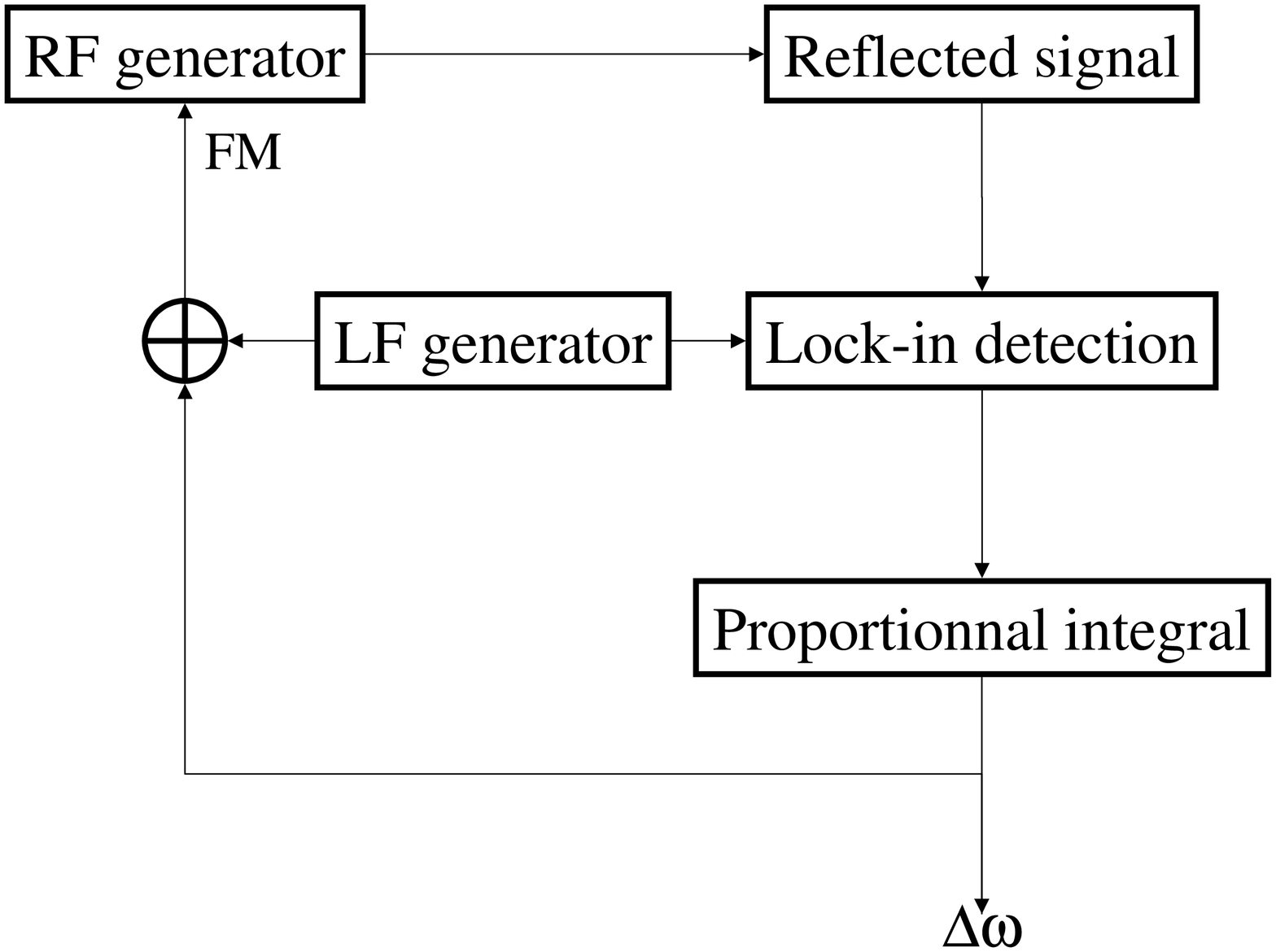}
    \end{center}
    \caption{Experimental setup used to lock the frequency of the RF 
    generator to the resonance frequency.}
    \label{fig:asservissement}
\end{figure}

The reflected signal of the resonator is measured with the setup of
figure \ref{fig:reflectedsignal} and used in a feedback loop to lock
the frequency of a RF generator to the resonance frequency.  The setup
is summarized in figure \ref{fig:asservissement}.  The resonator is
coupled capacitively to the external circuit using on-chip
capacitances. In order to preserve the quality factor of the resonator
we work in a configuration where the resonator is undercoupled.  The
RF power injected is sufficiently low ($\approx$ 10 pW) so as not to
heat the sample.

\subsubsection{Detection of the resonance frequency}
The frequency of the RF generator is modulated at $\Omega$ and the
signal from the resonator is detected by a lock-in detector at the
frequency of the modulation. The lock-in signal is to first
approximation the derivative of the resonance peak : it gives an error
signal i.e. this signal is zero at resonance, and changes sign when
the frequency of the generator is higher or lower than the resonance
frequency. Using this signal in a feedback loop the frequency of the
RF generator is locked to the resonance frequency. This way, by
measuring the feedback signal, one has directly access to the shift of
the resonance frequency. To enhance the accuracy we modulate the
magnetic field by a 1G AC field oscillating at 30 Hz, produced by a small 
superconducting coil close to the sample, and detect the modulated resonance 
frequency with a lock-in detector.

\subsubsection{Detection of the quality factor}
At this point we consider that the frequency of the generator is
locked to the resonance frequency by the previous setup.  The signal
measured is the signal reflected from the resonator. As a consequence
it is related to the reflexion coefficient
$(Z(\omega)-Z_0)/(Z(\omega)+Z_0)$, with $Z(\omega)$ the impedance of
the resonator and the coupling capacitance and $Z_0 = 50 \, \Omega$
the impedance matched by the external circuit. We assume that near the
resonance frequency the impedance $Z(\omega)$ reads~:
\begin{equation}
Z(\omega_0+\delta \omega)=\frac{R Q^2}{1+2 i Q \frac{\delta
\omega}{\omega_0}}
\end{equation} 
with $\omega_0$ the resonance frequency. In the limit $Z(\omega) \ll
Z_0$, which correspond to a very undercoupled resonator, the reflected
signal is a linear function of $Z(\omega)$. As a consequence if the RF
signal is frequency modulated at $\Omega$ around the resonance
frequency $\omega_0$ the reflected signal at $2 \Omega$ is related to
the second derivative of the real part $Z(\omega)$, which is
proportional to $Q^2$. This way by measuring the signal at $2 \Omega$
we have access to the quality factor.  However when the
frequency modulation is not small compared to the width of the
resonance peak or the resonator is not very undercoupled to the
external circuit, the relation between the signal at $2 \Omega$ and
the quality factor is not straightforward and needs calibration.

\section{Flux dependent polarisability}
\label{magnetopolarisabilty}
\begin{figure}[tb]
    \begin{center}
        \includegraphics[width=7.9cm]{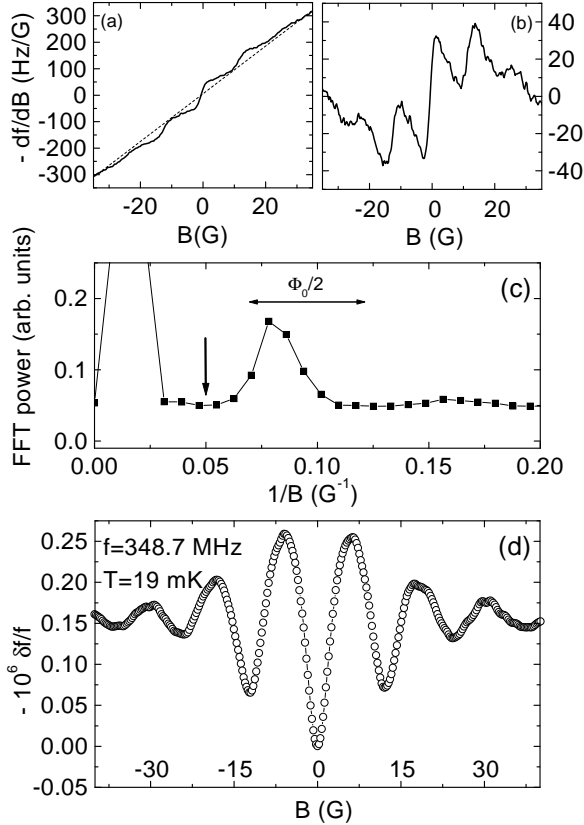}
    \end{center}
    \caption{(a) Derivative of the resonance frequency of the
      resonator with the rings versus magnetic field at illumination
    time 870 s. (b) Signal obtained by subtracting the base line (dashed line
    on graph (a)) due
    to the resonator from previous data.  (c) Fourier transform of
    signal (b). The vertical arrow indicates the cut-off frequency
    used for high pass filtering the signal.  (d) Frequency shift due
    to the rings obtained after integration of the high pass filtered
    signal of (b).}
    \label{fig:polarextract}
\end{figure}
In this part we present measurements of the flux dependent
polarisability of the rings, which are placed on the capacitive
part of the resonator as described in section
\ref{experimentalsetup}. In this configuration the resonance frequency
is decreased by 15 \%, due to the dielectric constant of the GaAs
substrate.  Moreover the quality factor drops down to 3000 at 20 mK at
zero illumination. This strong decrease is attributed to dielectric
losses in the heterojunction. The derivative of the resonance
frequency of the resonator with the rings is shown on figure
\ref{fig:polarextract} (a). This signal is a straight line, on top of
which small oscillations are superimposed. The straight line is due to
the field dependence of the penetration length in niobium, which
constitutes the resonator. This behaviour has been verified to be the
same with or without the rings.  The oscillating signal is on the
other hand attributed to the flux dependent electric response of the
rings. These oscillations are extracted by subtracting the base line (figure
\ref{fig:polarextract} (b)).  Note their anharmonicity as well as the 
existence of an aperiodic signal as illustrated by the Fourier
transform of the data (figure \ref{fig:polarextract} (c)) showing a well
defined peak. In order to focus on this periodic contribution, 
which is the expected signature of phase coherence, a
numerical high pass filter with a cut-off frequency of 0.05 G$^{-1}$
(corresponding to the arrow on figure \ref{fig:polarextract} (c)) is
applied and the signal is then numerically integrated in order to have
the frequency shift due to the rings (figure \ref{fig:polarextract}
(d)). This shift is proportional to the variation of polarisability
versus magnetic field according to formula \ref{frequencyshift}. We
will return to the aperiodic signal in the section devoted to
illumination effect.

\subsection{Magneto-polarisability}
The frequency shift due to the rings is periodic with a period of
approximately 12.5 G. From the Fourier transform (figure
\ref{fig:polarextract} (c)) the period of the oscillation is deduced
to be consistent with half a flux quantum $\Phi_0 / 2$ in a ring with
no signature of $\Phi_0$ periodicity, as expected for an 
Aharonov-Bohm effect averaged over many rings \cite{phi}. Note the extra 
broadening (by more than a factor 2) of this $\Phi_0/2$ peak compared 
to the measurements on a single connected ring. We interpret this 
as resulting from the dispersion in circumfrences in the different 
rings. The sign of frequency
shift is negative at low magnetic field which means according to
formula \ref{frequencyshift} that the magneto-polarisability is
positive , i.e.  $\alpha^{'}(H) - \alpha^{'}(0) > 0$ at low magnetic
field.  The screening is thus better when time reversal symmetry is
broken by magnetic field. The scale of the signal is given by the
amplitude of the first oscillation. From figure \ref{fig:polarextract}
(d) we deduced $\delta_{\Phi} f/f =(f(6.3 $\, G$)-f(0))/f = -2.5 \,
10^{-7}$. Note that this value means detecting a frequency shift of
100Hz on a frequency of 350 MHz.  With the coupling coefficient
estimated in appendix \ref{electric_coupling} it leads to the value
of the magnetopolarisability $\delta_\Phi \alpha^{'} / \alpha_{1D} = 5
\, 10^{-4} \pm 2.3 \, 10^{-4}$, where $\alpha_{1D}=\epsilon_0 \pi^2
R^3/\ln(R/W)$ is the calculated polarisability of a quasi one
dimensional (quasi-1D) circular ring of radius $R$.

\subsection{Theoretical predictions}
\begin{figure}[tb]
        \begin{center}
	        \includegraphics[width=7.9cm]{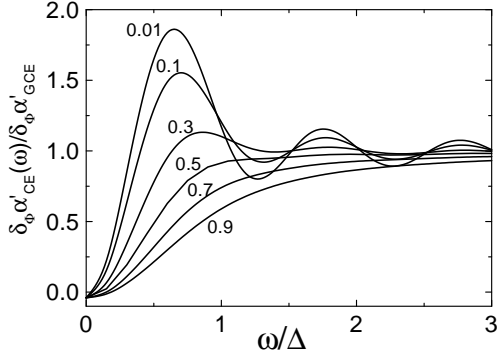}
        \end{center}
        \caption{Calculated evaluation of $\delta_{\Phi} \alpha'_{CE} 
(\omega)/\delta_{\Phi} \alpha'_{GCE}$ at different value of the
parameter $\gamma/\Delta$. Note that the value of
magnetopolarisability is zero at zero whatever the level broadening.}
        \label{fig:polarCE}
\end{figure}
Our experiment shows that there is a flux correction to the
polarisability of the rings, which is \textit{positive} at low
field. Let's now compare this result to recent theoretical
predictions. Since we are using a ring geometry we are going
alternatively from a situation where the system presents time reversal
symmetry (at flux values of $\Phi = n \Phi_0 / 2$, with $n \in
\mathbb{Z}$) to the case where time reversal symmetry is broken by
magnetic field. In the Random Matrix Theory (RMT) the first case
corresponds to Gaussian Orthogonal Ensemble (GOE) whereas the second
is related to the Gaussian Unitary Ensemble (GUE). So the quantity to
be compared with theoretical predictions, which evaluate the variation
of a physical variable $A$ between GOE and GUE, is $\delta_{\Phi} A$
defined as $A(\Phi_0/4)-A(0)$. Note that since the rings are
semi-ballistic, the transition with magnetic field may not be exactly
from GOE to GUE.

The polarisability of small metallic grains was studied using RMT
first by Gor'kov and Eliashberg \cite{gorkov65}. The sensitivity of
the electrostatic properties of mesoscopic systems to quantum
coherence has been emphasized by B{\"u}ttiker for connected geometries
\cite{buttiker}. The phase coherent correction to the polarisability
of isolated systems was recently theoretically investigated. Efetov
found that it is possible to relate self consistently this correction
to the flux dependence of the screened potential \cite{efetov96}. Two
recent works have calculated this effect in the diffusive regime using
linear response formalism (Noat {\it et al.} \cite{noat96,noat01}) or
supersymmetry techniques (Blanter and Mirlin
\cite{blanter98,blanter01}).

In the grand canonical ensemble (GCE) the chemical potential in each
ring is supposed to be constant. It describes a situation where the
rings are connected to a reservoir of particules.  \textit{A priori}
this is not the case in the experiment where the rings are isolated
but as the theory is simpler in GCE we recall first the result in this
statistical ensemble. No flux dependence for the polarisability is
predicted if the RF pulsation $\omega$ is much smaller than the
inverse relaxation time $\gamma$.  However when $\omega \gg \gamma$
the magnetopolarisability is related to the flux dependence of the
diagonal matrix element of the screened potential~:
\begin{equation}
\label{daGCE}
\delta_{\Phi} \alpha^{'}_{GCE} = - \frac{2 e^2}{E^2 \Delta}
\delta_{\Phi} \left( <|F_{\alpha \alpha}|^2>_{\mu} \right)
\end{equation}
$<|F_{\alpha \alpha}|^2>_{\mu}$ is the disorder averaged square of the
diagonal matrix element of the screened potential $F$ at energy $\mu$,
the mean chemical potential of the rings. $E$ is the applied electric
field. We note $\psi_{\alpha}$ the eigenstates of the unperturbed
system. This matrix element is then given by~:
\begin{multline}
\label{mpolar}
<|F_{\alpha \alpha}|^2>_{\mu}=\int d{\bf r_1} \int d{\bf r_2} F({\bf
r_1}) F({\bf r_2}) \\
<|\psi_{\alpha}({\bf r_1}) \psi_{\alpha}({\bf
r_2})|^2>_{\mu}
\end{multline}
From this expression it appears that the magnetopolarisability is
related to the difference of correlation function of the eigenstates
with and without time reversal symmetry. This correlation function has
been computed in the diffusive regime within a supersymetric
$\sigma$-model approach \cite{blanter96,mirlin01}~:
\begin{multline}
\label{eq:corrGOE}
V^2 <|\psi_{\alpha}({\bf r_1}) \psi_{\alpha}({\bf r_2})|^2>_{\mu}= \\
\left[ 1+2 k_d(r) \right] \left[ 1+2 \Pi_D({\bf r_1},{\bf r_2})
\right] \quad (GOE)
\end{multline}
\begin{multline}
\label{eq:corrGUE}
V^2 <|\psi_{\alpha}({\bf r_1}) \psi_{\alpha}({\bf r_2})|^2>_{\mu}= \\
\left[ 1 + k_d(r) \right] \left[ 1 + \Pi_D({\bf r_1},{\bf r_2})
\right] \quad (GUE)
\end{multline}
with $V$ the volume of the sample, $k_d(r)$ a short range function
which decays on the length scale of the mean free path and
$\Pi_D(\mathbf{r_1},\mathbf{r_2})$ the diffusion propagator. The
correction due to the short range term $k_d(r)$ has been shown to be
negligible \cite{blanter98}.  By considering only the diffusion term
the magnetopolarisability is given by~:
\begin{equation}
\label{polardiffuson}
\delta_{\Phi} \alpha^{'}_{GCE} = \frac{2 e^2}{E^2 \Delta V^2} \int
d{\bf r_1} \int d{\bf r_2} F({\bf r_1}) F({\bf r_2}) \Pi_D({\bf
r_1},{\bf r_2})
\end{equation}
Note that this derivation of the magnetopolarisability is equivalent
to the one used by Noat \textit{et al} \cite{noat96} based on the
following RMT argument~:
\begin{equation}
        \delta_{\Phi} \left( <|F_{\alpha \alpha}|^2> \right) \approx
-\frac{1}{2} <|F_{\alpha \alpha}|^2>_{GOE}
\end{equation}
This relation can also be obtained from \ref{eq:corrGOE} and
\ref{eq:corrGUE} using the fact that~:
\begin{equation}
        \int d{\bf r} F({\bf r}) = 0
\end{equation}
due to symmetry properties of the screened potential.  The calculation
of the magnetopolarisability using formula \ref{polardiffuson} for the
case of a quasi-1D ring (appendix \ref{quasi1Dpolar}) leads to~:
\begin{equation}
\frac{\delta_{\Phi} \alpha^{'}_{GCE}}{\alpha_{1D}} = \epsilon_r
f(\frac{L}{W}) \frac{\lambda_s}{W} \frac{\Delta}{E_c}
\label{da}
\end{equation}
$f(x)$ is a function related to the geometry and the dimension of the
sample.  Using this expression and the value of table
\ref{tab:CharacteristicsOfTheRings} we have $\delta \alpha^{'}_{GCE}
/\alpha_{1D} = 1.2 \, 10^{-3}$.

In our experiment the rings are isolated, the number of electrons in
each ring is supposed to be constant, so that the result of the
canonical ensemble (CE) should apply. At T=0 and zero frequency the
flux dependent correction to polarisability is found to be
zero. However at $\omega \gg \Delta$ the GCE result is recovered. The
complete frequency dependence of the magnetopolarisability in the CE
has been recently derived by Blanter and Mirlin
\cite{blanter01}. Following their reasoning but taking into account
the level broadening $\gamma$ we can write~:
\begin{equation}
        \delta_{\Phi} \alpha'_{CE} (\omega) = \delta_{\Phi}
        \alpha'_{GCE} F(\omega)
\end{equation}
with $F(\omega)$ a function which depends only on the statistic of
energy levels :
\begin{multline}
        F(\omega)=1+\int_0^{\infty} d\epsilon \frac{1}{\epsilon}
        \left( \frac{\epsilon
        (\epsilon+\omega)+\gamma^2}{(\epsilon+\omega)^2+\gamma^2}
        +\frac{\epsilon
        (\epsilon-\omega)+\gamma^2}{(\epsilon-\omega)^2+\gamma^2}
        \right)\\ \left[ \delta_{\Phi} R_2(\epsilon) +\int_0^\epsilon
        d\epsilon_1 \delta_{\Phi} R_3(\epsilon,\epsilon_1) \right]
\end{multline}
$R_2(\epsilon)$ and $R_3(\epsilon,\epsilon_1)$ are respectively the
two and three levels correlation function, known from RMT
\cite{mehta,blanter96bis}.  By evaluating this expression versus
frequency at different value of level broadening we get the results
shown on figure \ref{fig:polarCE}.

The behaviour at low value of the level broadening is in qualitative
agreement with result of reference \cite{blanter01}. In particular the
magnetopolarisability is found to be zero at zero frequency (in our
calculation the value of $\delta_{\Phi} \alpha'_{CE} (\omega=0)$ is at
least 25 time smaller than $\delta_{\Phi} \alpha'_{GCE}$). The present
experiment was performed at $\omega/\Delta = 0.2$, the CE
magnetopolarisability is equal at most to 50 \% of the GCE value (in
the limit of small level broadening). As a consequence the expected
value for $\delta_{\Phi} \alpha'_{CE}/\alpha_{1D}$ is then $6 \,
10^{-4}$ which is of the same order of magnitude as the experimental
value. Note that the measurement is not sufficiently accurate to give
an estimate of the level broadening by comparing the experimental
result with the curves of figure \ref{fig:polarCE}. A very interesting
extension of the experiment would be to study the
magnetopolarisability at different frequencies in order to test the
theoretical predictions. This could be done by working with resonators
with smaller inductances.

\subsection{effect of temperature}
\begin{figure}[tb]
    \begin{center}
        \includegraphics[width=7.9cm]{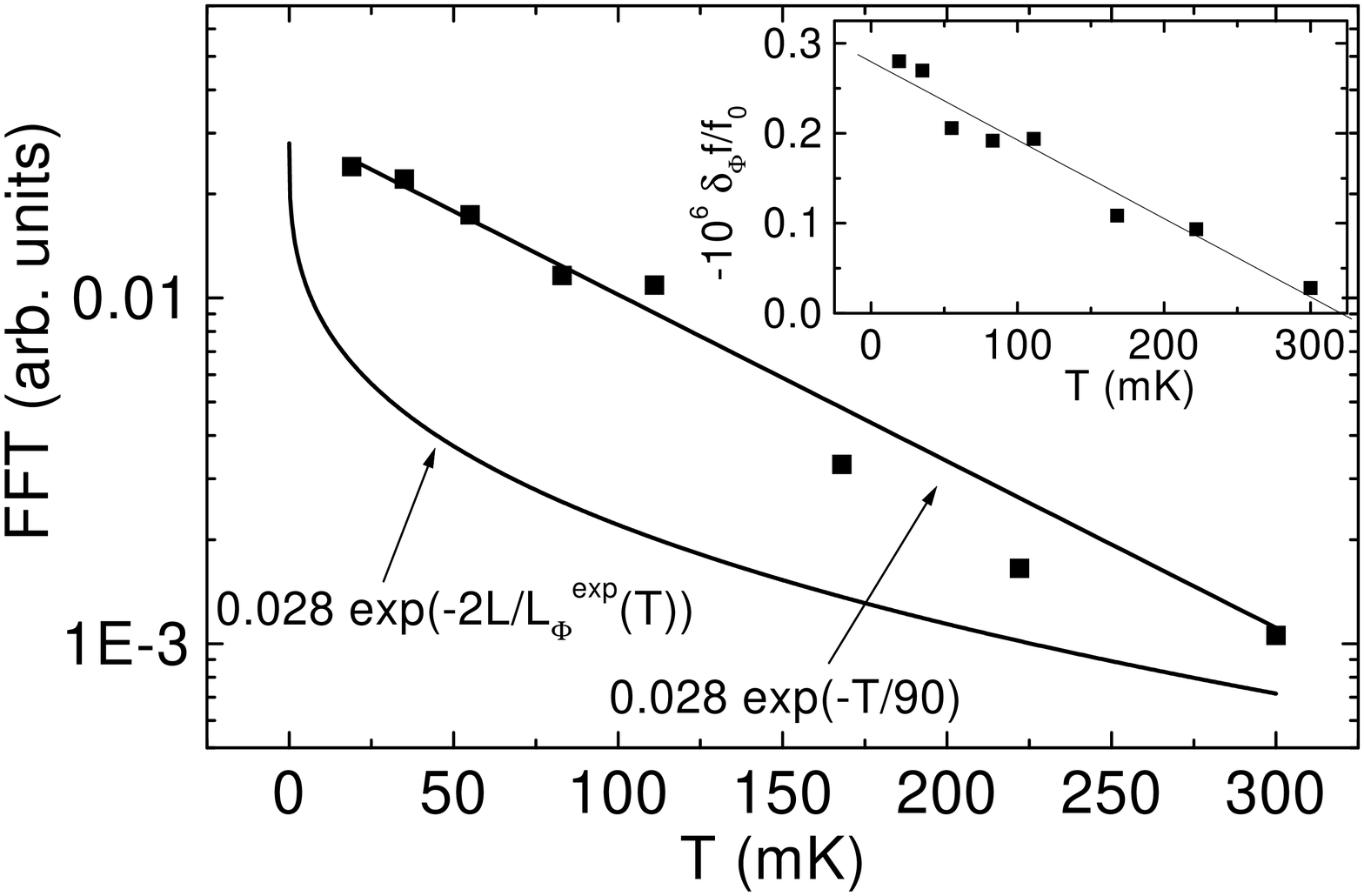}
    \end{center}
    \caption{Temperature dependence $\Phi_0/2$ component of the Fourier 
    transform of the signal. The fitting function used is proportional
    to $\exp{(- 2 L/L_{\Phi}(T))}$ with two fitting function for
    $L_{\Phi}(T)$.  First we took the phase coherence length measured
    on connected sample $L_{\Phi}^{exp}(T)$, which exhibits a
    $T^{-1/3}$ behaviour. The other fitting function is $L_{\Phi(T)}
    \propto 1/T$.  The best agreement is found with an exponential
    decay with a temperature scale of 90 mK. Inset : Temperature
    dependence of the frequency shift due to the rings.}
    \label{fig:polartemp}
\end{figure}
The temperature dependence of the signal is also investigated.  The
magnetopolarisability decreases with temperature (inset of figure
\ref{fig:polartemp}). Theoretically the effect of temperature on
magnetopolarisability has not been studied yet.  We will base our analysis
of the temperature dependence on the hypothesis that the amplitude of
the signal is related to the phase coherence length $L_{\Phi}$ in the
same way as weak-localization.  In this case the amplitude of the
$\Phi_0/2$ component of the signal is proportional to $\exp{(- 2
L/L_{\Phi}(T))}$ \cite{aronov87}. In figure \ref{fig:polartemp} the
temperature dependence of this component is shown. We have tried to
fit it using two laws for $L_{\Phi}(T)$.  First using the behaviour
deduced from the measurements on connected wires \cite{reulet95epl} we
have tried the experimental value $L_{\Phi}^{exp}(T)$ which exhibits a
$T^{-1/3}$ dependence, as expected for 1D system \cite{altshuler}.  It
leads to a poor agreement with experimental points. Using for the
phase coherence time the result of electron-electron interaction in
quantum dots (OD system) \cite{sivan94} $\tau_{\Phi}(T) \propto
T^{-2}$ , leading in the diffusive regime to $L_{\Phi}(T) \propto
1/T$, gives a better agreement. In this case the temperature scale is
found to be 90 mK. We deduced from this value
$\gamma=1/\tau_\Phi=D/L_\Phi^2 \simeq 0.8$ mK at 18 mK, i.e. much
smaller than the level spacing. The phase coherence length deduced from this analysis
is 10 times higher than the length measured on connected sample \cite{reulet95epl}.
We relate this difference between the non-connected
and connected case to the fact that whereas the connected samples
are one dimensional with a continuous energy spectrum due to the strong 
coupling with the reservoirs, the spectrum of the non-connected rings is discrete. 
As a concluding remark on this temperature dependance, we want to emphasize the need
of a deeper theoretical analysis of the behaviour of the magnetopolarisability versus
temperature.

\subsection{Effect of illumination}
\begin{figure}[tb]
        \begin{center}
                \includegraphics[width=7.9cm]{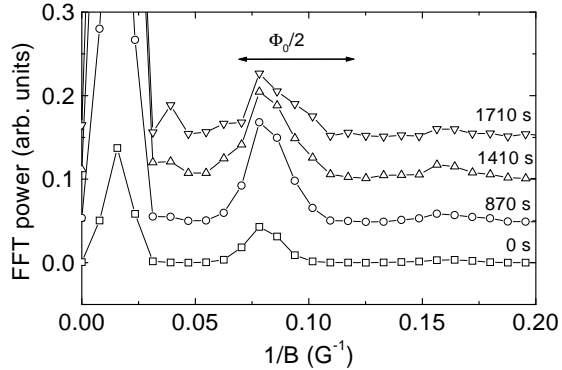}
        \end{center}
        \caption{Fourier transform of the derivative of the resonance 
frequency versus magnetic field at different illumination. The curves
are shifted for clarity.}
        \label{fig:FFTpolarillum}
\end{figure}
\begin{figure}[tb]
    \begin{center}
        \includegraphics[width=7.9cm]{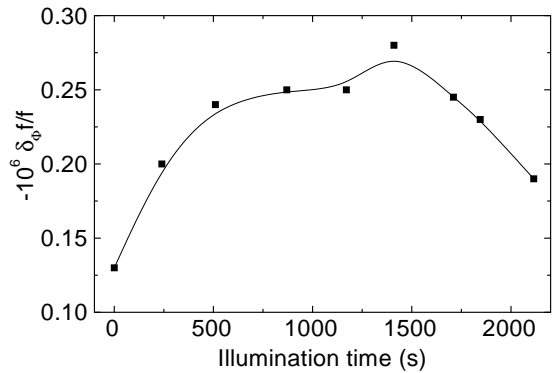}
    \end{center}
    \caption{Amplitude of the frequency shift due to the rings at different 
    illumination time.}
    \label{fig:effetillum}
\end{figure}
\begin{figure}[tb]
        \begin{center}
        	\includegraphics[width=7.9cm]{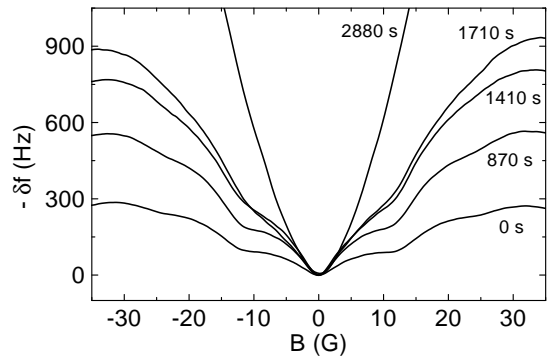}
        \end{center}
        \caption{Frequency shift without selecting the 
        $\Phi_0/2$ component before numerical integration, for different
        illumination times.}
        \label{fig:polarstep}
\end{figure}
Using the procedure described in \ref{sec:rings} we are able to study
the influence of electronic density on magnetopolarisability.  On
figure \ref{fig:FFTpolarillum} the Fourier transform of the derivative
of frequency shift, when the base line due to the resonator is removed,
is shown at different illumination time.  As expected the Fourier
transform exhibits a $\Phi_0/2$ peak.  Note also the low frequency
component which corresponds to the aperiodic signal seen on figure
\ref{fig:polarextract} (b).  The $\Phi_0/2$ peak depends on electronic
density. Its amplitude shows first an increase and then decreases at high
illumination. Moreover the width of the $\Phi_0/2$ peak increases
showing that the rings widen with illumination. The peak becomes
asymmetric suggesting that rings having long circumferences initially
not populated contribute to the signal at high illumination.
Following the procedure described in section
\ref{magnetopolarisabilty} we measured the amplitude of
magnetopolarisability versus illumination time. It yields figure
\ref{fig:effetillum}, which shows the change of the amplitude of the
magnetopolarisability.  At first the signal increases, then 
for illumination time above 1400 s the amplitude of oscillation decreases.
We interpret this behaviour in the following way. Before illumination
the electronic density in the rings obtained after deep etching of the
2DEG is strongly depressed compared to the nominal value. As a
consequence an important fraction of the rings are likely to be
localized and do not contribute to the magnetopolarisability.  In this
regime the signal is small. After illumination the number of rings
contributing to the signal increases so that the frequency shift
due to the rings increases. At high enough electronic density when the
rings are sufficiently populated so that they contain delocalized
electrons the theoretical results obtained in the diffusive regime 
are expected to be valid leading to a $1/g$ dependence (formula \ref{da}), 
with $g=E_c/\Delta$ the dimensionless conductance. This is a possible explanation 
for the decrease of the magnetopolarisability observed at high illumination
level. Note that we cannot exclude also a reduction of the screening
length $\lambda_s$ due to illumination. From formula \ref{da} we deduce 
then a decrease of the signal. However we believe that 
the screening length changes very weakly with illumination because this 
length is essentially determined by the density of states at the Fermi 
energy, which is independent of energy for a 2D system (see table 
\ref{tab:CharacteristicsOfTheRings}).

We have analysed so far the $\Phi_0/2$ periodic component of the
magnetopolarisability signal obtained after filtering low frequency
(see figure \ref{fig:polarextract}). On the other hand the whole
integrated signal is depicted in figure \ref{fig:polarstep}. One can
clearly see a triangular shape dependence of the signal with magnetic
field superimposed to the oscillations, very similar to the
weak-localisation conductance of the connected mesh shown in figure
\ref{fig:angr}. Note that this behaviour is only present at low
temperature, it completely disappears for temperature higher than 300
mK.  The amplitude of this extra signal due to the finite width of the
rings strongly increases and sharpens with illumination. We think that
it is reasonable to attribute this evolution to the increase of the
width of the rings. Note that a similar evolution has been previously
observed in the AC magnetoconductance of ballistic GaAs squares
\cite{noat98}.
 
\section{Electric absorption}
\begin{figure}[tb]
    \begin{center}
        \includegraphics[width=7.9cm]{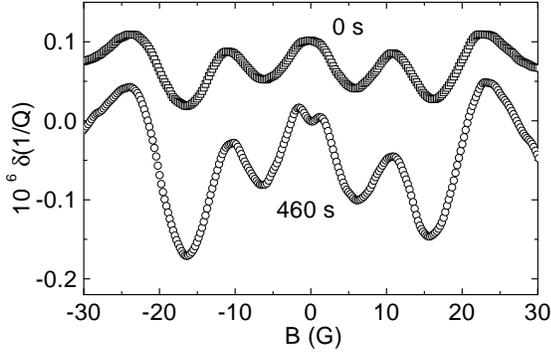}
    \end{center}
    \caption{Variation of 1/Q versus magnetic field at different illumination.}
    \label{fig:fig2}
\end{figure}
\begin{figure}[tb]
    \begin{center}
        \includegraphics[width=7.9cm]{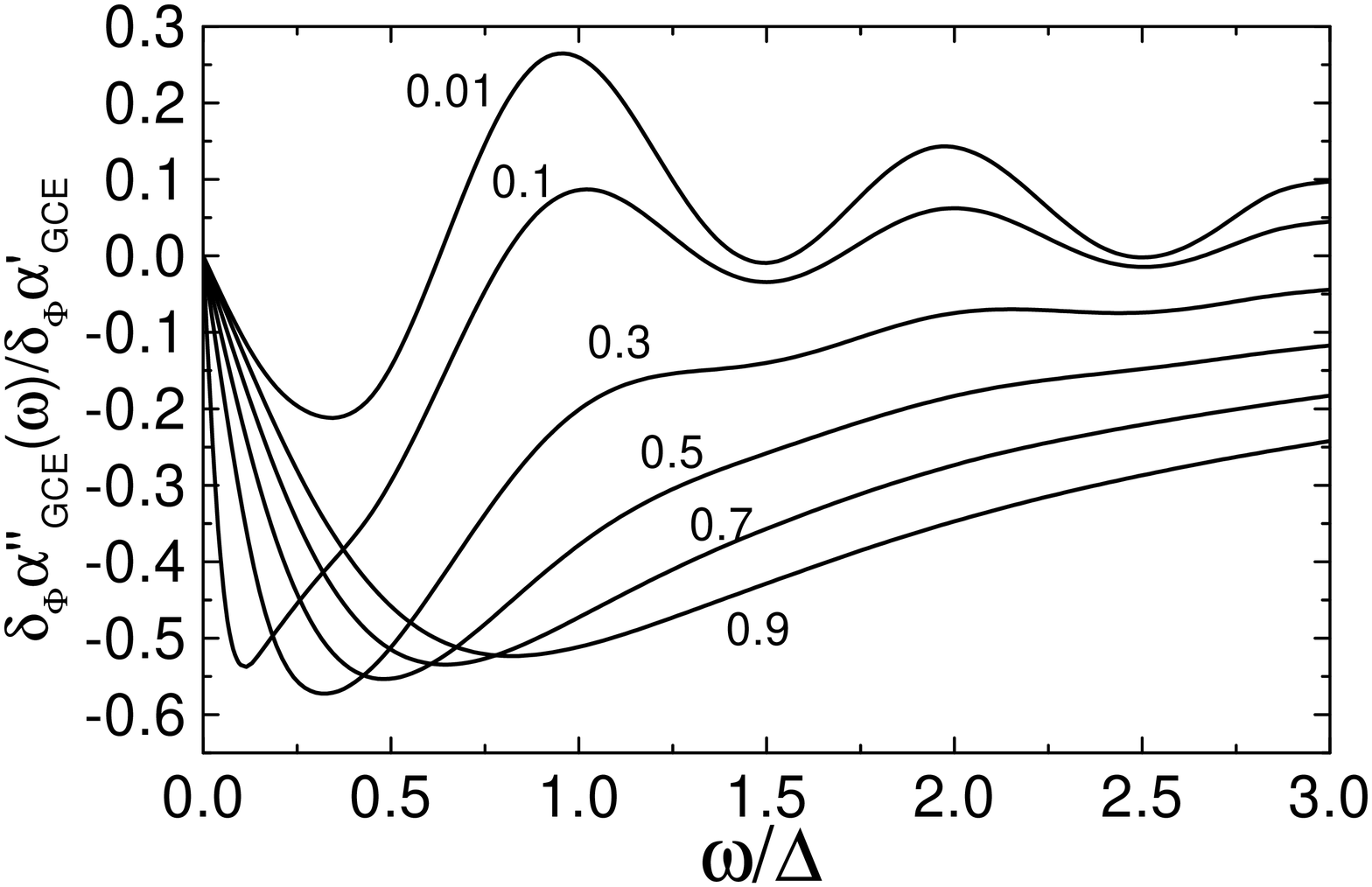}
    \end{center}
    \caption{Calculated value of $\delta_{\Phi}
      \alpha^{''}_{GCE}(\omega)/ \delta_{\Phi} \alpha^{'}_{GCE}$ 
      versus $\omega$ in GCE at different value of the parameter
      $\gamma/\Delta$.}
    \label{fig:electricabsorption}
\end{figure}
\begin{figure}[tb]
        \begin{center}
                \includegraphics[width=7.9cm]{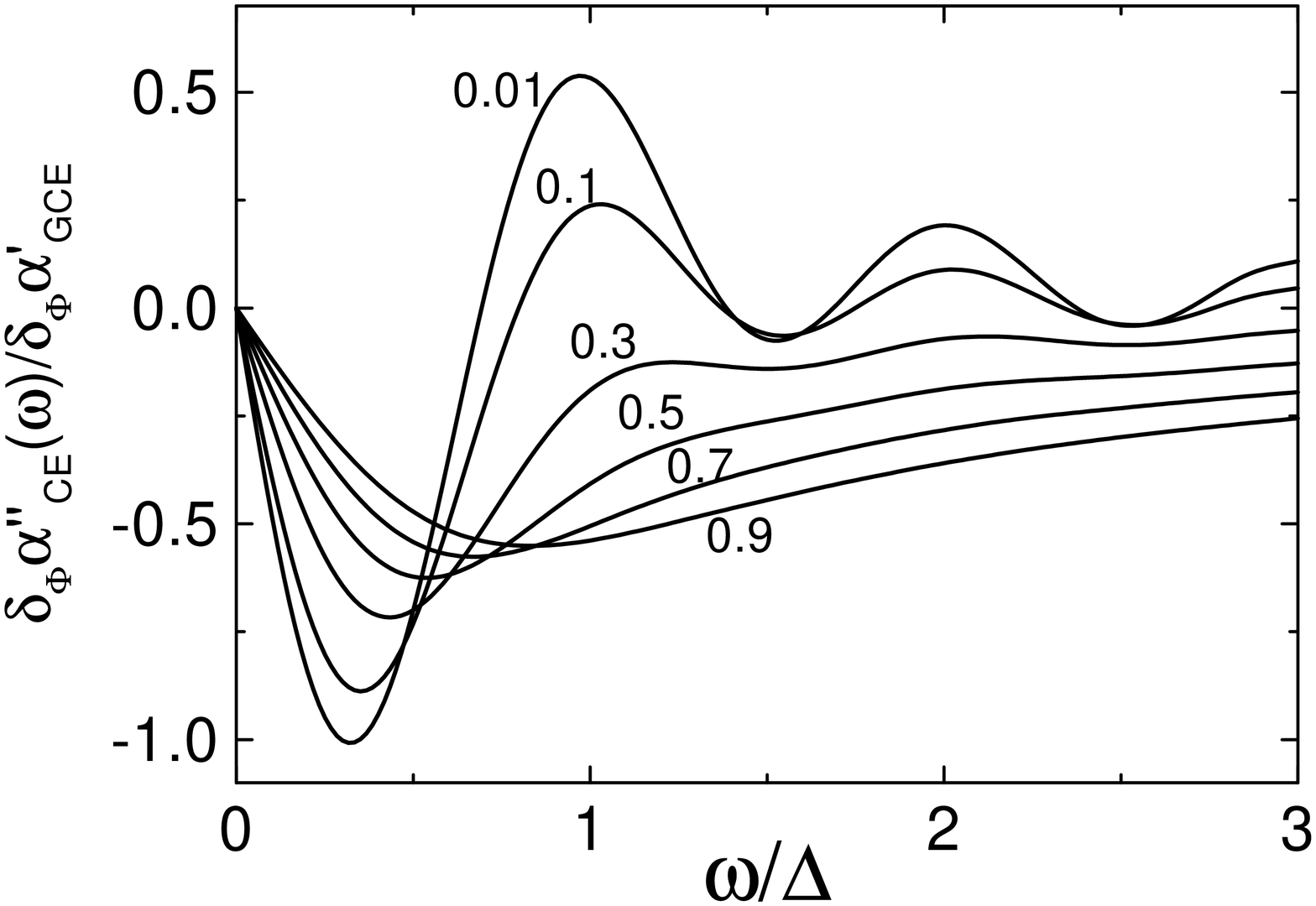}
        \end{center}
        \caption{Calculated electric absorption in CE versus frequency
          at different value of the parameter $\gamma/\Delta$}
        \label{fig:absorpCE}
\end{figure}By measuring the quality factor of the resonator versus magnetic
field, we have access to the flux dependent electric absorption
(formula \ref{Qshift}), which is related to the conductance 
, in the case of an electric dipole, through :
\begin{equation}
G_e = \frac{\omega \alpha^{''}}{a^2}
\end{equation} 
The contribution due to the rings (figure 
\ref{fig:fig2}) exhibits the same periodicity as the frequency shift, which 
corresponds to half a quantum flux in a ring. The low field signal decreases. 
It corresponds to a negative magnetoconductance, i.e. opposite to 
weak-localization. This surprising sign was pointed out in the context of the
magnetoconductance of rings submitted to an oscillating magnetic flux
in the discrete spectrum limit \cite{trivedi88,kamenev93,kamenev94}.

To explain this result one has to take into account the level spacing
distribution in a disordered system \cite{gorkov65,sivan87}. The sign
and amplitude of the typical variation of electric absorption are
understandable using the fact that level repulsion in a disordered
system is higher in GUE than in GOE. Following reference
\cite{noat98SM} the flux dependent electric absorption in a system
described by eigenvalues $\epsilon_\alpha$ and the corresponding
eigenfunctions $\psi_{\alpha}$ could be written in a linear response
regime~:
\begin{multline}
\label{absorption}
\delta_{\Phi} \alpha^{''} = - \frac{2 e^2}{E^2} \delta_{\Phi} \left(
\sum_{\alpha \neq \beta} \frac{f_{\alpha}-f_{\beta}}{\epsilon_{\alpha
\beta}} \frac{\gamma \omega}{(\epsilon_{\alpha
\beta}+\omega)^2+\gamma^2} |F_{\alpha \beta}|^2  \right. \\
\left. + \frac{\gamma \omega}{\gamma^2 + \omega^2} \sum_{\alpha} \frac{\partial
f_{\alpha}}{\partial \epsilon_{\alpha}} |F_{\alpha \alpha}|^2 \right)
\end{multline}
with $\epsilon_{\alpha \beta}=\epsilon_{\alpha}-\epsilon_{\beta}$.  We
will first consider this expression in the GCE where an average over
the chemical potential is computed. With this procedure
$<(f_{\alpha}-f_{\beta})/(\epsilon_{\alpha \beta})> =-1/\Delta \mu$
and $<\partial f_{\alpha}/\partial \epsilon_{\alpha}>= -1/\Delta \mu$,
where $\Delta \mu$ is the range over which the average over the
chemical potential is done. The first term in the right hand side of
equation \ref{absorption} then reads ~:
\begin{equation}
\label{nondiag}
\frac{2 e^2}{E^2 \Delta \mu} \delta_{\Phi} \left(
\sum_{\epsilon_{\alpha} \neq \epsilon_{\beta}} \frac{\gamma
\omega}{(\epsilon_{\alpha \beta}+\omega)^2+\gamma^2} |F_{\alpha
\beta}|^2 \right)
\end{equation}
Note that in this sum the energies $\epsilon_{\alpha}$ and
$\epsilon_{\beta}$ have to belong to the range $[\mu-\Delta \mu
/2,\mu+\Delta \mu /2]$. Using this constraint we replace the sum by an
integral~:
\begin{equation}
\sum_{\epsilon_{\alpha} < \epsilon_{\beta}} ( \, ) = \frac{\Delta
\mu}{\Delta} \int_0^{\infty} \frac{d\epsilon}{\Delta} R_2(\epsilon) (
\, )
\end{equation}
with $R_2(\epsilon)$ the two energy level correlation function.  In
this expression we neglect the flux dependence of the matrix element
and note its average value $<|F_{\alpha \beta}|^2>_{\mu,\omega}$.
With this approximation equation \ref{nondiag} reads~:
\begin{multline}
\frac{2 e^2}{E^2 \Delta} <|F_{\alpha \beta}|^2>_{\mu,\omega}
\int_0^{\infty} \frac{d\epsilon}{\Delta} \left( \frac{\gamma \omega}
{(\epsilon+\omega)^2+\gamma^2} \right. \\
\left. + \frac{\gamma
\omega} {(\epsilon-\omega)^2+\gamma^2} \right) \delta_{\Phi}
R_2(\epsilon)
\end{multline}
The Debye term of equation \ref{absorption} is equal in the GCE to~:
\begin{equation}
\frac{2 e^2}{E^2 \Delta} \frac{\gamma \omega}{\omega^2+\gamma^2}
\delta_{\Phi} \left( |F_{\alpha \alpha}|^2 \right)
\end{equation}
In the GCE in the dynamical regime the flux correction to the
polarisability is given by formula \ref{daGCE} at $T=0$. Using the
following correlation function \cite{mirlin01}~:
\begin{multline}
\label{corrnondiag}
V^2 <\psi_{\alpha}^{*}(\mathbf{r_1}) \psi_{\beta}(\mathbf{r_1})
\psi_{\alpha}(\mathbf{r_2})
\psi_{\beta}^{*}(\mathbf{r_2})>_{\mu,\omega}= k_d(r) \\
+ \left[1+k_d(r) \right] \Pi_D(\mathbf{r_1},\mathbf{r_2}) \quad (GOE)
\end{multline} 
we have $\delta_{\Phi} \left( <|F_{\alpha \alpha}|^2>_{\mu} \right)
\approx - <|F_{\alpha \beta}|^2>_{\mu,\omega}^{GOE}$. Hence :
\begin{multline}
\label{rapport_abs_pol}
\frac{\delta_{\Phi} \alpha^{''}_{GCE}(\omega)}{\delta_{\Phi}
\alpha^{'}_{GCE}} = -\frac{\gamma \omega}{\omega^2+\gamma^2}  +
\int_0^{\infty} \frac{d\epsilon}{\Delta} \left( \frac{\gamma \omega}
{(\epsilon+\omega)^2+\gamma^2} \right. \\
\left. + \frac{\gamma \omega}
{(\epsilon-\omega)^2+\gamma^2} \right) \delta_{\Phi} R_2(\epsilon)
\end{multline}
which can be evaluated numerically (figure
\ref{fig:electricabsorption}).

For isolated rings we have to apply the result of CE. It is then
possible to estimate the correction to electric absorption by using
the same treatment as for the real part of polarisability. It leads
to~:
\begin{multline}
        \frac{\delta_{\Phi} \alpha^{''}_{CE}(\omega)}{\delta_{\Phi}
        \alpha^{'}_{GCE}}=\int_0^{\infty} \frac{d\epsilon}{\Delta}
        \frac{1}{\epsilon} \left( \frac{\omega
        \gamma}{(\epsilon+\omega)^2+\gamma^2} \right. \\
        \left. +\frac{\omega\gamma}{(\epsilon-\omega)^2+\gamma^2} \right)
        \left[ \delta_{\Phi} R_2(\epsilon) +\int_0^\epsilon
        d\epsilon_1 \delta_{\Phi} R_3(\epsilon,\epsilon_1) \right]
\end{multline} 
Numerical estimation of this formula at different value of the level
broadening leads to the behaviour shown on figure
\ref{fig:absorpCE}. The electric absorption is always negative at low
frequency and may change sign at low value of $\gamma/\Delta$.

In order to compare these calculations with our experimental result we
will compute the ratio $\delta_{\Phi} \alpha^{''} / \delta_{\Phi}
\alpha^{'}$. It is worth noting that in our experiment this quantity
is given according to equations (\ref{frequencyshift},\ref{Qshift}) by
$\delta_{\Phi} (1/Q)/(-2 \delta_{\Phi} f/f)$ and is independent of the
number of rings coupled to the resonator or the electric coupling
coefficient $k_e$.  Experimentally we find $\delta_{\Phi}
\alpha^{''}/\delta_{\Phi} \alpha^{'} = -0.2$ at illumination time zero
and -0.23 after 420 s of illumination. Theoretically, at
$\omega/\Delta=0.2$, $\delta_{\Phi} \alpha^{'}_{CE}(\omega)= 0.5 \,
\delta_{\Phi} \alpha^{'}_{GCE}$ and $\delta_{\Phi}
\alpha^{''}(\omega)/\delta_{\Phi} \alpha^{'}_{GCE}=-0.5$ so that the
expected value of $\delta_{\Phi} \alpha^{''}(\omega)/\delta_{\Phi}
\alpha^{'}_{CE}(\omega)$ is around 1. It corresponds to small $\gamma/\Delta$. 
For higher value of this parameter the ratio
is of the same amplitude or higher. As a consequence the predicted
behaviour is consistent with the experimental value for the sign, but
the theoretical amplitude is too high by more than a factor 2. This
conclusion is different from our previous statement where the
frequency dependence of the real part of magnetopolarisability was not
taken into account \cite{deblock00}.

\section{Magnetic response}
\label{magneticresponse}
\begin{figure}[tb]
    \begin{center}
        \includegraphics[width=7.9cm]{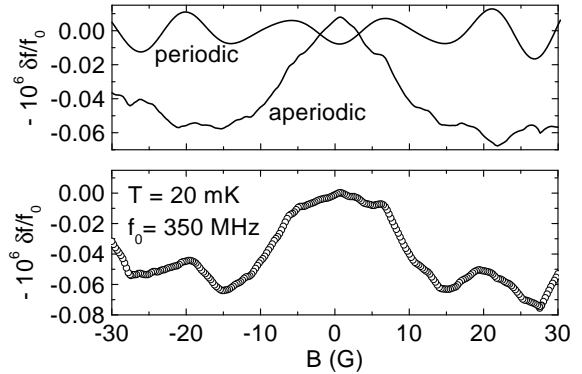}
    \end{center}
    \caption{Lower part : frequency shift due to the rings at 20 mK
      and zero illumination time. Upper part : the previous signal is
    decomposed into a periodic behaviour and a low frequency behaviour. }
    \label{fig:magneticsignal}
\end{figure}
\begin{figure}[tb]
    \begin{center}
        \includegraphics[width=7.9cm]{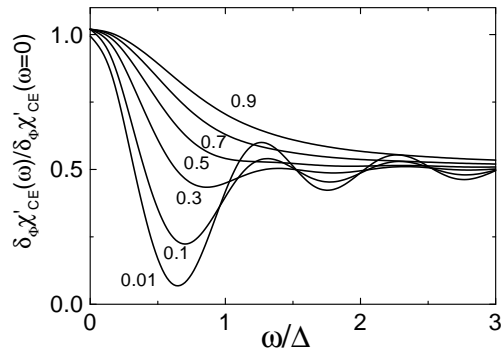}
    \end{center}
    \caption{Calculated frequency dependence of the real part of the 
susceptibility at different value of the parameter $\gamma/\Delta$.}
    \label{fig:magneticCE}
\end{figure}
\begin{figure}[tb]
    \begin{center}
        \includegraphics[width=7.9cm]{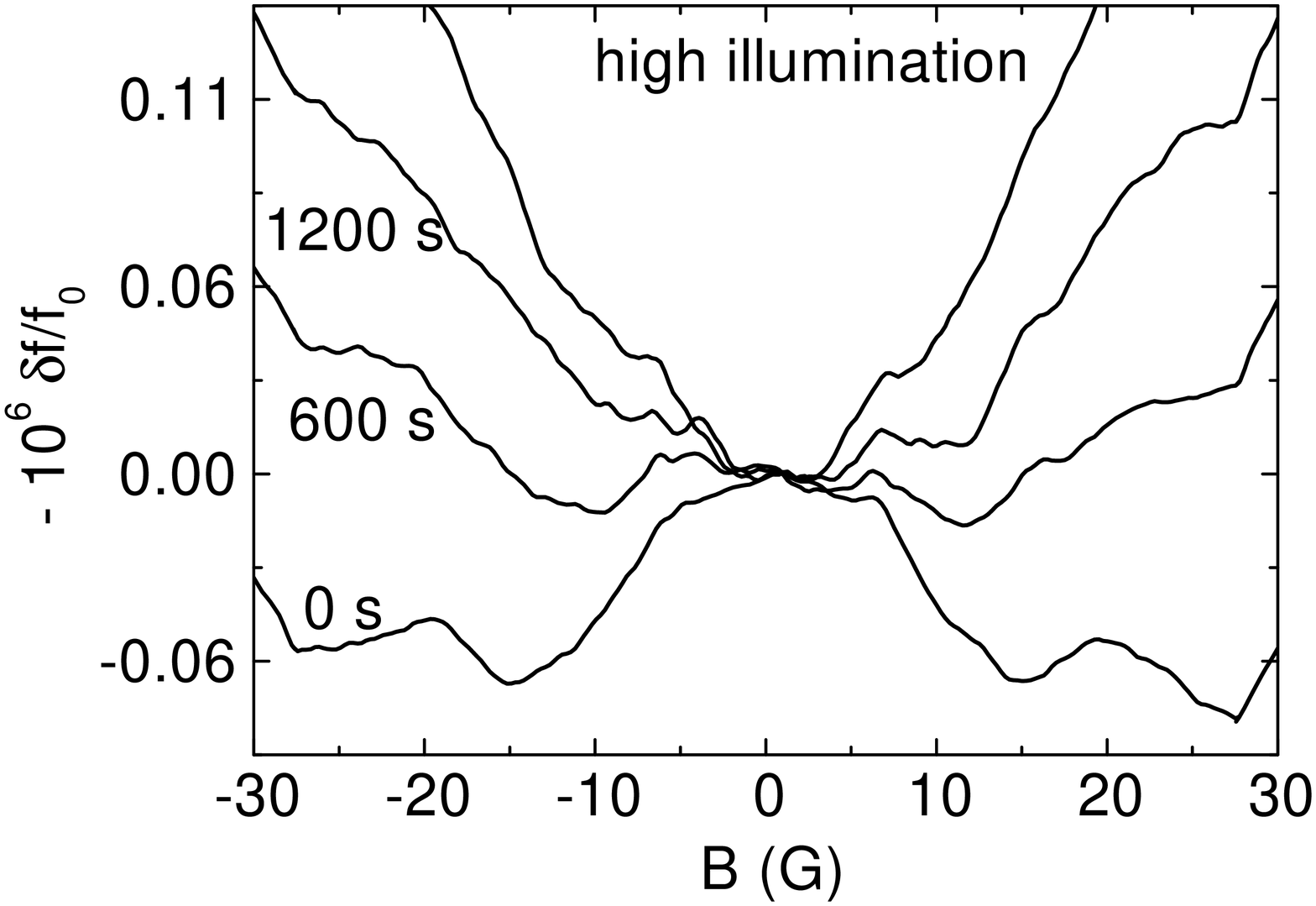}
    \end{center}
    \caption{Frequency shift due to the magnetic response of the rings 
at different illumination time.}
    \label{fig:magneticillum}
\end{figure}
\begin{figure}[tb]
    \begin{center}
        \includegraphics[width=7.9cm]{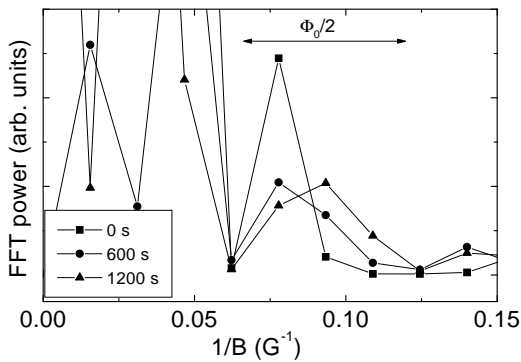}
    \end{center}
    \caption{FFT of the magnetic response of the rings at 
    different illumination.}
    \label{fig:magneticillumFFT}
\end{figure}
Due to the design of the resonator we have also the opportunity to
investigate the magnetic response of the same Aharonov-Bohm rings used
for the measurements of the electric response. In this case the rings
are placed on top of the inductive part of the resonator. Note that to do so 
we have to warm-up, cool down and re-illuminate the rings. As a consequence,
strictly speaking, the rings are not the same than for the measurement of the 
electrical response because the electronic density and the disorder realisation
in each ring is not exactly the same from one run to the other. Nevertheless 
due to the fact that we are measuring an ensemble average quantity the change 
in disorder realisation of each ring does not modify the result of the experiment.
Moreover we have checked (on the electric response measurement) that, for the same 
illumination procedure, the result varies within a 15 \% range from one run 
to the other. The flux dependent orbital magnetism at a frequency of 350 MHz is then
detected.  In this configuration the quality factor of the resonator
is only 500. We do not understand this strong increase of magnetic
losses. This low quality factor decreases the accuracy of our
measurements of the resonance frequency. Moreover it prevents precise
measurements of the flux dependence of the dissipative part of
magnetic response of the rings. As a consequence in this part we
present only the flux dependent non dissipative part of the magnetic
response. Note that we cannot rule out the fact that the signal
measured in this configuration could be partially due to electric
response of the rings.  However due to the small value of the residual
capacitance of the meander line and the bad electric coupling in this
geometry this electric component is estimated to be at least 20 times
smaller than when the rings are placed on top of the
capacitance. Moreover the very different shape of the electric and
magnetic signals is a strong evidence that we are indeed measuring
essentially the magnetic response of the rings.

\subsection{Flux dependent orbital magnetism}
The signal measured at zero illumination, after subtracting the
base line due to the resonator, is shown in the lower part of figure
\ref{fig:magneticsignal}. Inspired by our previous analysis we
decompose the measured field dependent part of the signal into an
aperiodic and a periodic part which corresponds to $\Phi_0/2$ in a
ring (figure \ref{fig:magneticsignal}).
We interpret the $\Phi_0/2$ component as the contribution of
electronic trajectories enclosing the whole ring. On the other hand
the triangular shape signal could be due to the contribution of
trajectories confined in the finite width of the ring. The amplitude
of the $\Phi_0/2$ periodic component of the signal is 
$\delta_{\Phi}f/f=-1.5 \, 10^{-8}$. We deduce from formula \ref{dfmagnetic} and
evaluation of the magnetic coupling done in appendix
\ref{magnetic_coupling} that the flux dependent magnetic response of
the ring is $\delta_{\Phi} \chi = 5.4 \, 10^{-24} \pm 2.1 \, 10^{-24}$
m$^{-3}$. In the following we first assume that the main contribution
to this signal is due to the flux derivative of the persistent
currents \cite{reulet94} and then discuss finite frequency effects.
If the flux dependence of persistent currents is $I(\Phi) = I_0
\sin{(4 \pi \Phi/\Phi_0)}$, we deduce~:
\begin{equation}
I_0 = - \frac{\delta_{\Phi} \chi}{2 \mu_0} \frac{\Phi_0}{4 \pi S^2}
\end{equation}
with $S$ the surface of the ring. We find then a \textit{diamagnetic}
average persistent current, the amplitude of which is : $|I_0|=0.25
\pm 0.1$ nA. The aperiodic component of the signal corresponds on the
other hand to low field paramagnetism.

\subsection{Persistent currents}
Let's now compare our result for the average persistent currents to
other experimental results and to theoretical predictions. A $\Phi_0/2$
periodic diamagnetic persistent currents has been also observed in
arrays of metallic rings \cite{jariwala01,levy90}. The expected
amplitude of the averaged current due to repulsive electron-electron
interactions from first order Hartree-Fock calculation \cite{ambegaokar90} is 
$E_c/\Phi_0 = 1.5$ nA, this value is expected to be decreased by higher order
terms. Considering on the other hand theoretical predictions
for non interacting electrons \cite{altshuler91} the expected value is
between $\sqrt{\Delta E_c}/\Phi_0 = 0.6$ nA and $\Delta/\Phi_0 = 0.3$
nA. In both cases the currents are predicted to be paramagnetic. The
rather small difference between interacting and non interacting
electrons is very specific to the GaAs rings where the number of
electrons is small. The measured signal is consistent for the
amplitude but not for the sign (unless assuming attractive
interactions) with theoretical predictions.

It may also be important to take into account an effect of frequency
for the flux dependent orbital magnetism. In fact by applying a
formalism very similar to the one used for magnetopolarisability the
variation of the real part (non-dissipative) of the susceptibility of
a ring submitted to an oscillating magnetic flux in CE without
interactions is given by \cite{trivedi88,reulet94}~:
\begin{equation}
        \delta_{\Phi} \chi^{'}(\omega) = \delta_{\Phi} \left(
        \sum_{\alpha \neq \beta}
        \frac{f_{\alpha}-f_{\beta}}{\epsilon_{\alpha \beta}} 
         \frac{\epsilon_{ \alpha \beta}(\epsilon_{\alpha
        \beta}+\omega)+\gamma^2}{(\epsilon_{\alpha
        \beta}+\omega)^2+\gamma^2} |J_{\alpha \beta}|^2 \right)
\end{equation}
with $J_{\alpha \beta}$ the matrix element of the current operator. It
is then possible to apply the same reasoning as for the real part of
polarisability, and to use the fact that $\delta_{\Phi} (<|J_{\alpha
\alpha}|^2>)=<|J_{\alpha \beta}|^2>$, so that~:
\begin{multline}
        \frac{\delta_{\Phi} \chi^{'}(\omega)}{\delta_{\Phi}
        \chi^{'}(\omega=0)}=\frac{1}{2} \left( 1-\int_0^{\infty}
        d\epsilon \frac{1}{\epsilon} \right. \\ \left. \left( \frac{\epsilon
        (\epsilon+\omega)+\gamma^2}{(\epsilon+\omega)^2+\gamma^2}
        +\frac{\epsilon
        (\epsilon-\omega)+\gamma^2}{(\epsilon-\omega)^2+\gamma^2}
        \right) \right. \\ \left. \left[ \delta_{\Phi} R_2(\epsilon) 
        +\int_0^\epsilon
        d\epsilon_1 \delta_{\Phi} R_3(\epsilon,\epsilon_1) \right]
        \right)
\end{multline} 
The evaluation of this expression is easily deduced from the
evaluation of magnetopolarisability and leads to figure
\ref{fig:magneticCE}. Frequency induces a strong decrease of the magnetic
signal for frequencies of the order of $\Delta$ but does not seem to induce a sign
change of $\delta_{\Phi} \chi'$. Note that in strong contrast with the
electric response the magnetic susceptibility is maximum at zero
frequency, which corresponds to persistent currents.  It would be
important to investigate the effect of finite frequency on the
contribution due to electron-electron interactions on persistent
currents.

Recently it has been suggested that the measured currents may be due to a
rectifying behaviour of the rings : a high frequency noise leads 
then to a DC current \cite{kravtsov93}. Noise also induces dephasing. A
recent paper by Kravtsov and Altshuler \cite{kravtsov00} predicts that
those two quantities, average persistent current and dephasing
measured by the saturation of the phase coherence time, are related in
a simple way $I = C e / \tau_{\Phi}(T=0)$.  $C$ is a constant giving
the sign of the persistent current and $\tau_{\Phi}(T=0)$ the
dephasing time at zero temperature.  Using the value of
$\tau_{\Phi}=1.5$ ns at 20 mK, deduced from measurements on connected
sample, and considering the orthogonal case (absences of spin orbit,
then $C=-4/\pi$) we deduce an expected value for persistent currents
of -0.14 nA. The predicted persistent current is then
\textit{diamagnetic}. The sign and amplitude are then consistent with
our experimental findings. On the other hand if we take the value
deduced from the temperature dependance of the magnetopolarisability
of non-connected rings, which is not the case considered
theoretically, we deduce $I_0=- 0.02$ nA, smaller by an order of
magnitude than the experimental value.

\subsection{Effect of illumination}
The influence of electronic density on the magnetic response of the
rings has been investigated by illuminating them. Different
illumination times are shown in figure \ref{fig:magneticillum}. We
observed that the triangular envelope of the signal changes sign and
increases with illumination.
For each illumination the Fourier transform of the signal exhibits a
component which is consistent with half a flux quantum periodicity
(figure \ref{fig:magneticillumFFT}). One sees however that with
illumination the shape of the $\Phi_0/2$ peak in the Fourier transform is
modified. The peak broadens with electronic density indicating that
the width of the rings increases. Note that this width is always
consistent with the one deduced from etching and depletion effects. The
power of the Fourier transform integrated in the $\Phi_0/2$ zone is
constant within 10 \%. So the amplitude of the $\Phi_0/2$ signal is
constant but its shape is modified.  It indicates that the amplitude
of the persistent currents does not depend much on electronic density.
The sign change of the low frequency part of the flux dependent
magnetic response of the rings, going from low field paramagnetism to
diamagnetism, is not understood.

\section{Conclusion}

We have presented measurements of electric and magnetic responses of
Aharonov-Bohm rings etched in a 2DEG. They present a flux dependent 
correction to screening.  This
correction is positive in low field which means that screening is
enhanced when time reversal symmetry is broken by a magnetic
field. The sign of the effect is consistent with theory for isolated
rings at finite frequency. The value of magnetopolarisability is
$\delta_\Phi \alpha^{'} / \alpha_{1D} = 5 \, 10^{-4} \pm 2.3
\,10^{-4}$, with $\alpha_{1D}=\epsilon_0 \pi^2 R^3/\ln(R/W)$ the
calculated polarisability of a quasi-1D circular ring of radius R. 
The temperature dependence of
magnetopolarisability is consistent with $\L_{\Phi} \propto 1/T$. The
behaviour versus electronic density is compatible with a $1/g$
dependence of magnetopolarisability.

The magnetic response has been measured on the very same array of rings.  
The rings exhibits a signal consistent with \textit{diamagnetic} average
persistent currents of amplitude $|I_0|=0.25 \pm 0.1$ nA.

Because the measurements are done on the same rings it is possible to
compare the electric and the magnetic signal. The experimental ratio
between the frequency shift due to the electric or magnetic response
is around 10, a value consistent with theoretical expectations
taking into account the electric and magnetic coupling 
coefficient (appendix \ref{magnetic_coupling} and \ref{electric_coupling})
and the ratio between the typical matrix element of the 
screened potential and the current operator which leads to \cite{noat98SM} :
\begin{equation}
\label{eq:rapelmagn}
	\frac{\delta_{\Phi} \chi}{\delta_{\Phi} \alpha/\epsilon_0} 
	\approx \left( Z_0 G_D \right)^2 \approx {\mathcal{\alpha}}^2 g^2 
\end{equation}
with $Z_0 = \sqrt{\mu_0/\epsilon_0} = 377 \Omega$ the vacuum impedance, 
$G_D = g e^2/h$ the Drude conductance and $\alpha \approx 1/137$ the fine 
structure constant. We have thus shown that the mesoscopic electromagnetic 
response of GaAs rings is dominated by the flux dependent polarisability 
instead of orbital magnetism. This would not be the case in metallic 
rings, where, due to the very short screening length, the mesoscopic electric 
response is negligible. The low field diamagnetic sign of the orbital 
magnetism needs further investigation both on the experimental and theoretical 
sides.

\section{Acknowledgments}
We thank B. Etienne for the fabrication of the heterojunction. We
acknowledge fruitful discussions with L.P. L\'evy, Ya.M. Blanter, 
S. Gu\'eron and G. Montambaux and technical help of M. Nardone 
and P. Demianozuck.

\appendix

\section{Evaluation of the magnetic coupling}
\label{magnetic_coupling}
 
In this appendix we evaluate the magnetic coupling of one square ring
with the resonator in the configuration of the experiment. The
inductance $\mathcal{L}$ is modelled by two cylindrical wires
separated by a distance of $2 d$ (see figure \ref{fig:res_anneaux}
(a)). A ring is submitted to the magnetic field of those wires. Let's
first evaluate the mutual inductance $\mathcal{M}$ between the ring
and the resonator. Using Ampere theorem the magnetic field generated
by a current $I$ in the the inductance is easily calculated. The flux
of this magnetic field through a ring of size $a$ located at the (0,0)
point is then~:
\begin{equation}
    \Phi = {\mathcal{M}} I = \frac{\mu_0 a}{\pi} \ln \left( \frac{2 d
    + a}{2 d -a} \right) I
\end{equation}
If the ring is located at a point (x,y) $\mathcal{M}$ reads~:
\begin{equation}
    {\mathcal{M}} = \frac{\mu_0 a}{4 \pi} \ln \frac{
    ((x+d+\frac{a}{2})^2+y^2)((x-d-\frac{a}{2})^2+y^2)}{((x+d-\frac{a}{2})^2
    +y^2)((x-d+\frac{a}{2})^2+y^2)}
\end{equation}
The ring submitted to a magnetic field $B$ acts as a magnetic dipole
$m = \chi B/ \mu_0$. This dipole is equivalent to the ring with a
current $m / a^2$, so that the flux in the inductance is now $\Phi =
({\mathcal{L}} + {\mathcal{M}}^2 \chi/a^4) I$. We deduce from these
results that $k_m = {\mathcal{M}}^2/(\mu_0 a^4 {\mathcal{L}})$ with
${\mathcal{L}}$ the inductance of the meander-line. From the resonance
frequency and the calculation of the capacitance we deduce
${\mathcal{L}} = 0.05 \mu$H, .  The rings are not perfectly well
coupled to the inductance so that they are not all located at
$x=0$. Moreover because of the mylar sheet inserted between the rings
and the resonator the rings are not in the plane of the resonator. To
take this into account the inductance is averaged over the x position
of the rings and $1.5 \mu$m$ < y < 2.5 \mu$m. Within this
approximations : $k_m = 1.3 \, 10^{11} \pm 0.5 \, 10^{11}$ m$^{-3}$.

\section{Evaluation of the electric coupling}
\label{electric_coupling} 
\begin{figure}[tb]
    \begin{center}
        \includegraphics[width=8cm]{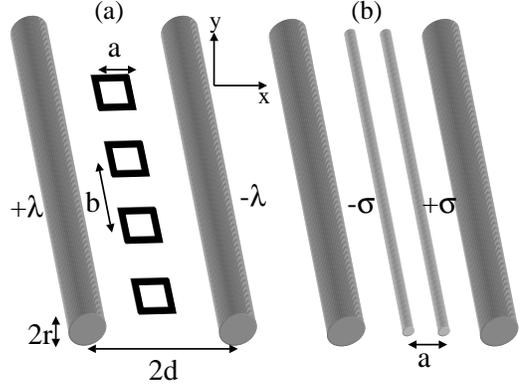}
    \end{center}
    \caption{a: Schematic picture of the rings coupled to the
    capacitance. b: Modelisation used for the estimation of the
    electric coupling coefficient. The lineic charge $\sigma$ is
    determined by the polarisability of the rings and the electric
    field generated by the capacitance at the ring position. }
    \label{fig:res_anneaux}
\end{figure}
In this appendix we evaluate the electric coupling coefficient $k$ of
one ring with the capacitance $\mathcal{C}$ of the resonator.  The
capacitance is modelled by two cylindrical wires of radius $r$ and
separated by a distance $2 d$, one wire with a linear charge of
$\lambda$, the other one $-\lambda$. The electric field outside the
wires is the one generated by two lines of lineic charge $\lambda$ and
$-\lambda$ separated by a distance $2 d_1$ determined by $d_1 =
\sqrt{d^2-r^2}$ \cite{landau}.  In our case ${\mathcal{C}} = 10$ pF.
Using Gauss theorem we can easily calculate the electric field in the
plane of the rings in every point $(x,y)$ outside the wires~:
\begin{equation}
    E(x,y) = \frac{\lambda}{2 \pi \epsilon_0} \left(
    \frac{x+d_1}{(x+d_1)^2+y^2}-\frac{x-d_1}{(x-d_1)^2+y^2} \right)
\end{equation}
A ring submitted to this field generates an electric dipole $P =
\alpha E$ with $\alpha$ the polarisability of one ring, so that the
rings submitted to electric field act as an ensemble of dipole. We
model them by two line of lineic charge $\sigma$ and $-\sigma$
separated by a distance $a$, and such that $\sigma b = \alpha E / a$
(see figure \ref{fig:res_anneaux}). Note that to do so the electric
field has to be constant on the scale of the rings : this is roughly
the case.  Evaluating the potential $\delta V$ created by these two
wires between each side of the capacitance, and using the relation
$\delta V = - V \delta \mathcal{C} / \mathcal{C}$ we have for rings
located at (x,y)~:
\begin{equation}
    \frac{\delta \mathcal{C}}{\mathcal{C}} = \frac{\sigma}{2 \lambda}
    \frac{\ln \frac{ ((d
    -r-x-\frac{a}{2})^2+y^2)((d-r+x-\frac{a}{2})^2+y^2)}
    {((d-r-x+\frac{a}{2})^2+y^2)((d-r+x+\frac{a}{2})^2+y^2)}} {\ln
    \frac{d_1^2}{r^2}}
\end{equation}
To have the capacitance shift induced by one ring we have to divide
the previous result by the number of rings $N=l/b$ with $l$ the length
of the capacitance. Moreover as the ring are imbedded in GaAs-AlGaAs
we have to divide our result by the dielectric constant of the
substrate $\epsilon_r = 12.85$. We can now evaluate the electric
coupling coefficient $k_e$ defined by $\delta {\mathcal{C}} /
{\mathcal{C}} = N k_e \alpha$ by averaging over the x-position of the
rings and considering that the rings are located between 1.5 $\mu$m
and 2.5 $\mu$m in the y-direction from the resonator. Within these
approximations $\epsilon_0 \epsilon_r k_e = 8 \, 10^{10} \pm 3.4 \,
10^{10}$ m$^{-3}$.  Note that the previous result is very close to the
value of the magnetic coupling coefficient $k_m$.

\section{Magnetopolarisability for a quasi-1D ring}
\label{quasi1Dpolar}

In this part we evaluate the magnetopolarisability given by formula
\ref{polardiffuson} for a quasi-1D ring. The diffusion propagator at
frequency $\omega$ is given by~:
\begin{equation}
        \Pi_D(\mathbf{r},\mathbf{r'},\omega)=\frac{\Delta S}{\pi}
        \sum_n \frac{\psi_n^{*}(\mathbf{r}) \psi_n(\mathbf{r'})}{-i
        \omega + E_n}
\end{equation}
$S$ and $\Delta$ are respectively the surface of the ring and the mean
level spacing.  $E_n$ and $\psi_n$ are the eigenvalues and
eigenvectors of the diffusion equation~:
\begin{equation}
        - \hbar D \Delta_r \psi_n(\mathbf{r}) = E_n \psi_n(\mathbf{r})
\end{equation}
We consider a 2D ring of perimeter $L$, radius $R$ and width $W$, with
$W \ll L$.  In this case the solutions of the diffusion equation are~:
\begin{equation}
        \psi_{m,n}(x,y)=\sqrt{\frac{2}{L W}} \cos{\pi m \frac{y}{W}}
        \exp{i 2 \pi n \frac{x}{L}}
\end{equation}
with $m \in \mathbb{N^{*}}$ and $n \in \mathbb{Z}$. The modes
corresponding to $m=0$ are given by~:
\begin{equation}
        \psi_{m=0,n}(x,y)=\sqrt{\frac{1}{L W}} \exp{i 2 \pi n
        \frac{x}{L}}
\end{equation}
$x$ is the coordinate along the ring, and $y$ the radial
coordinate. In our description the ring corresponds to $y \in
[0,W]$. We consider reflecting border in the $y$ direction.The
corresponding eigenvalue is~:
\begin{equation}
        E_{m,n} = E_c \left[ 2 \pi n^2 + m^2 \frac{\pi}{2} \left(
        \frac{L}{W} \right)^2 \right]
\end{equation}
$E_c = h D / L^2$ is the Thouless energy.  The mean charge density
(average over the width of the ring) in the ring submitted to an
electric field $E$ is~:
\begin{equation}
        \rho(x =R \cos{\theta},y)=\frac{\epsilon_0 \pi R E}{W
        \ln{(R/W)}} \cos{\theta}
\end{equation}
Note that using this density we recover the classical result for a
quasi-1D ring $\alpha_{1D}=\epsilon_0 \pi^2 R^3/\ln(R/W)$. In the
Thomas-Fermi approximation we deduce the mean screened potential~:
\begin{equation}
        F(x = R \cos{\theta},y)=\frac{R \lambda_s E}{2 W \ln{(R/W)}}
        \cos{\theta}
\end{equation}
Using this relation and the formula for the diffuson one can do the
calculation analytically. Because of the form of $F$ only the mode
(m=0,n=1) remains and leads to~:
\begin{equation}
\frac{\delta_{\Phi} \alpha^{'}}{\alpha_{1D}} = \epsilon_r
f(\frac{L}{W}) \frac{\lambda_s}{W} \frac{\Delta}{E_c}
\end{equation}
with $f(x)=1/(4 \pi^2 \ln{x/2 \pi})$.


\begin{thebibliography}{}
\bibitem{wl} These values are deduced from weak localization
measurements on connected wires etched in the same heterojunction than
the rings.
\end{thebibliography}
\end{document}